\documentclass[lettersize,onecolumn]{IEEEtran}
\usepackage{amsmath,amsfonts}
\usepackage{algorithm}
\usepackage{array}
\usepackage[caption=false,font=normalsize,labelfont=sf,textfont=sf]{subfig}
\usepackage{textcomp}
\usepackage{stfloats}
\usepackage{url}
\usepackage{verbatim}
\usepackage{graphicx}
\usepackage{cite}
\newtheorem{theorem}{Theorem}

\newtheorem{lemma}{Lemma}
\newtheorem{definition}{Definition}
\newtheorem{proposition}{Proposition}
\newtheorem{remark}{Remark}
\newtheorem{proof}{Proof}
\usepackage{amssymb}
\usepackage{graphicx}
\usepackage{xcolor}
\usepackage{multirow}
\usepackage{graphbox}
\usepackage{paralist} 
\usepackage{algpseudocode}
\usepackage{xurl}
\usepackage{enumerate}  
\usepackage{float}
\usepackage{bm}

\usepackage{algpseudocode}
\usepackage{algorithmicx}
\usepackage{lipsum} 
\usepackage{tabulary}

\begin{document}

\title{Extended Gabidulin-Kronecker Product Codes and Their Application to Cryptosystems}

\author{
    \IEEEauthorblockN{Zhe Sun\textsuperscript{1}, Terry Shue Chien Lau\textsuperscript{2}, Mengying Zhao\textsuperscript{1,*},  Zimeng Zhou\textsuperscript{3}, Fang-Wei Fu\textsuperscript{4}}\\
    \IEEEauthorblockA{
        \textsuperscript{1}School of Computer Science and Technology, Shandong University, Qingdao 266100, China. \\
        \textsuperscript{2}{Faculty of Computing \& Informatics, Multimedia University, Persiaran Multimedia, 63100 Cyberjaya, Selangor, Malaysia.}\\
        \textsuperscript{3} The Key Laboratory of Dependable Service Computing in Cyber Physical Society, Ministry of Education of China, Chongqing University, Chongqing 400044, China.\\
        \textsuperscript{4}Chern Institute of Mathematics and LPMC, Nankai University, Tianjin 300071, China. \\
        Email: sunzhe@sdu.edu.cn, terry.lau@mmu.edu.my, zhaomengying@sdu.edu.cn, zhouzimeng@sdu.edu.cn, fwfu@nankai.edu.cn.
    }
    \thanks{*Corresponding author.}
}




\maketitle

\begin{abstract}
In this paper, we initiate the study of Extended Gabidulin codes with a Kronecker product structure and propose three enhanced variants of the Rank Quasi-Cyclic (RQC) (Melchor et.al., IEEE IT, 2018) cryptosystem. First, we establish precise bounds on the minimum rank distance of Gabidulin-Kronecker product codes under two distinct parameter regimes. Specifically, when $n_{1}=k_{1}$ and $n_{2}=m<n_{1}n_{2}$, the minimum rank distance is exactly $n_{2}-k_{2}+1$. This yields a new family of Maximum Rank Distance (MRD) codes, which are distinct from classical Gabidulin codes. For the case of $k_{1}\leq n_{1},k_{2}\leq n_{2},n_{1}n_{2}\leq m$, the minimum rank distance $d$ of Gabidulin-Kronecker product codes satisfies a tight upper and lower bound, i.e., $n_{2}-k_{2}+1 \leq d \leq (n_{1}-k_{1}+1)(n_{2}-k_{2}+1)$. Second, we introduce a new class of decodable rank-metric codes, namely Extended Gabidulin-Kronecker product (EGK) codes, which generalize the structure of Gabidulin-Kronecker product (GK) codes. We also propose a decoding  algorithm that directly retrieves the codeword without recovering the error vector, thus improving efficiency. This algorithm achieves zero decoding failure probability when the error weight is within its correction capability. Third, we propose three enhanced variants of the RQC cryptosystem based on EGK codes, each offering a distinct trade-off between security and efficiency. For 128-bit security, all variants achieve significant reductions in public key size compared to the Multi-UR-AG (Bidoux et.al., IEEE IT, 2024) while ensuring zero decryption failure probability--a key security advantage over many existing rank-based schemes. Our first variant, RQC.EGK-BWE, is based on EGK codes and blockwise errors. It achieves a public key size of 3949 bytes, which is approximately 4\% smaller than that of the Multi-UR-AG. The second variant, RQC.EGK-Multi-NH, is built on EGK codes and non-homogeneous errors. It achieves a public key size of 3679 bytes, representing an approximate 11\% reduction compared to the Multi-UR-AG. The third variant, RQC.EGK-Multi-UR, which is based on EGK codes, non-homogeneous errors, and unstructured, can provide enhanced security. It achieves a public key size of 2138, achieving an approximate 48\% improvement over the Multi-UR-AG.
\end{abstract}


\section{Introduction}
\IEEEPARstart{R}{ank} metric codes were first systematically introduced by Delsarte \cite{ref36} in 1978 in the context of bilinear forms over finite fields. The core innovation of this approach consists of replacing the Hamming weight with the rank weight, thereby establishing a new metric space and coding framework for error correction. Within the rank metric space, Gabidulin codes, as typical MRD codes, strictly satisfy the Singleton bound under the rank metric and thus exhibit optimal distance performance. This makes them the core theoretical support and building blocks for rank-metric code-based cryptosystems. In 1991, Gabidulin, Paramonov, and Tretjakov proposed the GPT cryptosystem \cite{ref25}, the first encryption scheme based on Gabidulin codes in the rank metric. Subsequently, Gibson and Overbeck conducted cryptanalysis \cite{ref38,ref39,ref40} on this cryptographic scheme and proved its insecurity. Consequently, a series of revised schemes have been proposed \cite{ref41,ref42,ref43,ref44,ref45,ref8,ref9,ref11,ref13,ref14,ref37} to resist these attacks. Code-based cryptography has garnered widespread attention from both industry and academia, and has emerged as one of the most promising candidates in post-quantum cryptography. In the ongoing NIST post-quantum cryptography standardization process, schemes based on Hamming metric codes, such as Classic McEliece, HQC, and BIKE, have progressed to the final rounds, and HQC has been selected as the final standard. In contrast, rank metric code-based cryptography, which offers potential advantages in computational efficiency and bandwidth, has been encouraged by NIST PQC for further investigation. 

The Rank Quasi-Cyclic (RQC) scheme was selected as a second-round candidate in the NIST post-quantum cryptography standardization process due to its competitive parameters. However, algebraic attacks \cite{ref18,ref19} required subsequent parameter adjustments to maintain security. To mitigate algebraic attacks without drastically increasing parameters, the concept of non-homogeneous errors was introduced \cite{ref7}. In this model, an error vector of length $n$ is divided into three parts, with specific constraints on the rank supports of these parts, where two parts have identical rank supports and the rank support of these two parts is smaller than that of the third part. This method allows reducing the decoding of $[2n,n]$ codes to the decoding of random $[3n,n]$ codes in the LRPC cryptosystem within the security reduction of RQC. In addition, Reference \cite{ref7} first proposed the Non-Homogeneous RSD problem (NHRSD), which is also the first method to reduce the parameters of RQC. Meanwhile, they also introduced Augmented Gabidulin (AG) codes (a new class of decodable codes that leverage the concept of support erasure for the rank metric) for use in the public decoding process. A year later, Reference \cite{ref10} proposed another method to improve RQC, and first introduced blockwise errors, the block decoding problem (BRD), and $l$-LRPC codes. The idea of blockwise errors is to partition an error of length $n$ into $l$ parts with pairwise disjoint rank supports, each part having a rank weight of $r_{i}$ and satisfying $\sum_{i=1}^{l}r_{i} = r$ (where $r$ denotes the total rank weight of the error). This method enables the achievement of a more compact key size. The aforementioned schemes employ LRPC codes as public codes, which are utilized in the public decryption process. Unlike the aforementioned schemes, Reference \cite{song2025interleaved} proposed the use of Extended Gabidulin codes as public codes, and further improved the efficiency of solving the block decoding problem (BRD). The proposed improved RQC scheme features a more compact key size.

\subsection{Motivations}
 Due to their optimal distance properties, Gabidulin codes have become foundational building blocks for rank-based cryptosystems. A prominent example is the Rank Quasi-Cyclic (RQC) cryptosystem \cite{ref6}, which was selected as a second-round candidate in the NIST post-quantum cryptography standardization process due to its compact key size and computational efficiency. However, this scheme failed to advance to the candidate list of the third round of NIST's post-quantum cryptography standardization process due to potential security issues. Nevertheless, its unique advantageous of compact size renders it still worthy of research. In recent years, many researchers have attempted to improve this scheme, such as in \cite{song2025interleaved,ref7,ref10}. Among them, Reference \cite{ref10} proposes to improve RQC by using the blockwise decoding problem and $l$-LRPC codes, yet the scheme has a non-zero decryption failure probability, which potentially increases security risks. Reference \cite{ref7} introduces non-homogeneous errors and AG codes to enhance RQC, but there remains a non-zero decryption failure probability. Later, Reference \cite{song2025interleaved} proposes the use of Extended Gabidulin codes and applies them to the RQC scheme. Unfortunately, the improved scheme still suffers from a non-zero decryption failure probability. Therefore, designing an RQC variant that maintains a compact key size while achieving zero decryption failure probability is a critical research objective, as decryption failures can lead to severe security vulnerabilities.

One promising avenue to design an RQC variant with both compact keys and zero decryption failure is through code composition techniques. The Kronecker product, in particular, provides a structured method to construct high-dimensional codes from lower-dimensional components while preserving desirable algebraic properties. The core advantage of the Kronecker product lies in the following. When generating a high dimensional matrix through the product operation of low-dimensional matrices, key algebraic properties of the high-dimensional matrix, such as rank characteristics and invertibility, can be directly derived from the low-dimensional matrices involved in the operation. This provides solid theoretical support for the controllable design of code structural performance. For instance, scholars such as Sun constructed a high-dimensional product code by combining two low-dimensional Gabidulin codes via Kronecker product operation, which initially verified the potential of this method in reducing key size \cite{sun2024}. Notably, the application of the Kronecker product to enhance the RQC scheme remains unexplored. This gap motivates our work.

\subsection{Contributions}
In this paper, we initiate using the Kronecker product structure to improve the RQC scheme, and design three secure and efficient cryptosystems based on the rank metric. Our contributions are threefold.

First, we present a refined analysis of Gabidulin-Kronecker product (GK) codes \cite{sun2024}, establishing tighter bounds on their minimum rank distance under two parameter regimes.

(1) When $n_{1}=k_{1}$ and $n_{2}=m<n_{1}n_{2}$, the minimum rank distance is exactly $n_{2}-k_{2}+1$, yielding a new family of MRD codes distinct from classical Gabidulin codes.

(2) When $k_{1}\leq n_{1},k_{2}\leq n_{2},n_{1}n_{2}\leq m$, the minimum distance $d$  satisfies $n_{2}-k_{2}+1 \leq d \leq (n_{1}-k_{1}+1)(n_{2}-k_{2}+1)$. Furthermore, experimental results have verified that both its upper and lower bounds are tight.

Second, we introduce a new class of rank-metric code, namely Extended Gabidulin-Kronecker product codes, and develop an efficient decoding algorithm for them. Unlike other existing decoding methods, the proposed algorithm does not require recovering the error vector but directly retrieves the codeword, which reduces the complexity. In addition, under some given conditions, compared with existing schemes \cite{song2025interleaved,ref7} that suffer from non-zero decryption failure probability, the proposed codes enable zero decryption failure probability.

Third, we leverage EGK codes to construct three enhanced variants of the RQC cryptosystem: (i) RQC.EGK-BWE (using blockwise errors), (ii) RQC.EGK-Multi-NH (using non-homogeneous errors), and (iii) RQC.EGK-Multi-UR (incorporating both non-homogeneous errors and an unstructured design). A key advantage of all three variants is their deterministic decryption (zero failure probability), alongside significant reductions in key size compared to the Multi-UR-AG and other contemporary candidates. Detailed parameter sets and performance comparisons in Section \ref{sec111} demonstrate that, for 128-bit security, all variants achieve significant reductions in public key size compared to the Multi-UR-AG. Specifically, the public key of our first scheme is 3949 bytes, which represents an approximate 4\% reduction compared to Multi-UR-AG scheme. The public key of our second scheme is 3679, which is 11\% smaller than that of the Multi-UR-AG scheme. The public key of our third scheme is 2138 bytes, which is 48\% smaller than that of the Multi-UR-AG scheme. This third variant also yields a more compact public key than the candidate schemes Classic McEliece, HQC, and RQC submitted to the fourth round of the NIST standardization process, especially at medium to high security levels.

\subsection{Organization}
The remainder of this paper is organized as follows. Section \ref{sec222} provides necessary background, including notations, preliminaries in coding theory and rank metric, the hard problems in coding theory, the public key encryptions and the key encapsulation mechanisms, the IND-CPA and IND-CCA2, and a review of the classic RQC cryptosystem. In Section \ref{sec3333}, we provide a detailed description of the minimum rank distance of Gabidulin-Kronecker product codes. In Section \ref{sec444}, we propose a new rank metric code, namely Extended Gabidulin-Kronecker product codes, and give the decoding algorithm of Extended Gabidulin-Kronecker product codes in Section \ref{sec555}. In Section \ref{sec666}, We apply our new proposed rank metric code to cryptography and propose three improved schemes for the RQC cryptosystem. We present the security analysis and the recommended parameters of our proposed cryptosystem in Section \ref{sec777}. Section \ref{sec888} sums up this paper.

\section{Preliminaries}
\label{sec222}
\subsection{Notations}
For clarity, we list the key notations used throughout this paper in Table \ref{tab:example1}.

\begin{table}[h]
\centering
\caption{Definitions of notations}
\label{tab:example1}
\begin{tabular}{|c|c|}
\hline
Notation & Definition  \\
\hline
$\mathbb{N}$ & The set of natural numbers \\       \hline
$m$ & A positive integer \\  \hline
$p$&  A prime \\ \hline
$q$& A power of a prime $p$ \\ \hline
$\mathbb{F}_{q}$ & The finite field with $q$ elements, where $q$ is a prime  \\ \hline
$\mathbb{F}_{q^m}$& The extension field of $\mathbb{F}_{q}$ of degree $m$\\ \hline
$\mathbb{F}_{q^m}^{n}$& A row vector of length $n$ over the field extension $\mathbb{F}_{q^m}$\\ \hline
$\mathbb{F}_{q}[X]$  &  The univariate polynomial ring with variable $x$ over the finite field $\mathbb{F}_{q}$\\
\hline
$GL_{n}(\mathbb{F}_{q})$   & The set of all $n \times n$ invertible matrices with coefficients over $\mathbb{F}_{q}$ \\ \hline
$x \stackrel{\$}{\leftarrow} \mathcal{S}$ &   The $x$ is selected uniformly and randomly from the finite set $\mathcal{S}$\\
\hline
$\mathcal{M}_{k,n}(\mathbb{F}_{q^m})$  &  The set of all $k \times n$  matrices with entries over the extension field $\mathbb{F}_{q^m}$ \\ \hline
PPT & An algorithm is a PPT algorithm if it is a probabilistic polynomial-time algorithm\\ \hline
$\mathcal{L}_{\leq k-1}$ & The set of all $q$-polynomials whose non-zero leading term has a $q$-degree less than or equal to $k-1$ \\ \hline
$\omega$ & The exponent of matrix multiplication with $2 \leq \omega \leq 3$ and a practical value is 2.81 when more than a few hundreds rows and columns \\ \hline
$\binom{a}{b}$ & Combination number, i.e., the number of ways to choose $b$ elements from $a$ elements to form a group \\ \hline
$\mathcal{O}(\cdot)$& Time complexity of algorithms in computer science \\ \hline
\end{tabular}
\end{table}

\subsection{Coding Theory and Rank Metric}
In this paper, all vectors are assumed to be row vectors. Let $\bm{z} = (z_1,\ldots,z_n) \in \mathbb{F}_{q^m}^{n}$. A vector $\bm{b} = (b_{1},\ldots,b_{m}) \in \mathbb{F}_{q^m}^{m}$ is called a basis vector of $\mathbb{F}_{q^m}$ over $\mathbb{F}_{q}$ if its entries form a basis for $\mathbb{F}_{q^m}$ as an $\mathbb{F}_{q}$-vector space. An element $\beta \in \mathbb{F}_{q^m}$ is called a normal element if $(\beta,\beta^{q},\ldots,\beta^{q^{m-1}})$ forms such a basis vector. 
\begin{definition}
    ($\mathbb{F}_{q^m}$-linear Codes) An $\mathbb{F}_{q^m}$-linear code $\mathcal{C}$ of length $n$ and dimension $k$ is a $k$-dimensional $\mathbb{F}_{q^m}$-linear subspace of $\mathbb{F}_{q^m}^{n}$. Its parameters are denoted by $[n,k]_{q^m}$.
\end{definition}

 Given an $[n,k]_{q^m}$ linear code $\mathcal{C}$, it can be represented by a generator matrix or equivalently a parity-check matrix. Here, a matrix $G$ is called a generator matrix of the code $\mathcal{C}$ if it satisfies: 
 $$\mathcal{C}=\{\bm{mG}:\bm{m}\in \mathbb{F}_{q^m}^{k}\}.$$
 
 A matrix $H$ is called a parity-check matrix of the code $\mathcal{C}$ if it satisfies: 
 $$\mathcal{C}=\{\bm{x}\in \mathbb{F}_{q^m}^{n}:H\bm{x}^{T}=\bm{0}\}.$$
 
 We say that the matrix $G$ is in systematic form if it has the form $[I_{k}|A]$, we say that the matrix $H$ is in systematic form if it has the form $[-A^{T}|I_{n-k}]$.

\begin{definition}
    (Rank Weight) Let $\{b_{1},\ldots,b_{m}\}$ be a fixed basis of $\mathbb{F}_{q^m}$ over $\mathbb{F}_{q}$. For a vector $\bm{z}=(z_{1},z_{2},\ldots,z_{n}) \in \mathbb{F}_{q^m}^{n}$, each component can be uniquely expressed as $z_{j} = \sum_{i=1}^{m}z_{ij}b_{i}$ with coordinates $z_{ij} \in \mathbb{F}_{q}$. The rank weight $wt_{R}(\bm{z})$ is defined as the rank of the associated coordinate matrix $M(\bm{z}) = (z_{i,j})_{1 \leq i\leq m, 1 \leq j \leq n} \in \mathbb{F}_{q}^{m \times n}$ over $\mathbb{F}_{q}$.
\end{definition}
\begin{definition} (Support)
   The support $Supp(\bm{z})$ of a vector $\bm{z}=\{z_{1},z_{2},\ldots,z_{n}\} \in \mathbb{F}_{q^m}^{n}$ is the $\mathbb{F}_{q}$-linear subspace of $\mathbb{F}_{q^m}$ spanned by its components $\{z_1,z_2,\ldots,z_n\}$. Equivalently, the rank weight of $\bm{z}$ can be regarded as the dimension of $Supp(\bm{z})$ over $\mathbb{F}_{q}$, i.e., $wt_{R}(\bm{z}) = dim(supp(\bm{z}))$. 
\end{definition}

  \begin{definition}
      (Rank Distance) For any two words $\bm{z}, \bm{y} \in$ $\mathbb{F}_{q^m}^{n}$, the rank distance between $\bm{z}$ and $\bm{y}$ is defined as $d_{R}(\bm{z},\bm{y}) = wt_{R}(\bm{z}-\bm{y})$. 
  \end{definition}
  \begin{definition}
      (Minimum Rank Distance) For a linear code $\mathcal{C}$, the minimum rank distance of $\mathcal{C}$ is $$d_{R}(\mathcal{C}) = min\{d_{R}(\bm{c}_{1},\bm{c}_{2})|\\ \bm{c}_{1},\bm{c}_{2}\in \mathcal{C}, \bm{c}_{1} \neq \bm{c}_{2}\}= min\{wt_{R}(\bm{c})|\bm{c} \in \mathcal{C}, \bm{c} \neq \bm{0}\}.$$ 
  \end{definition}

We now introduce ideal codes. To do so, we first recall some necessary background.

Let $N(X)\in \mathbb{F}_{q}[X]$ be a polynomial of degree $n$, and let $\phi=\mathbb{F}_{q^m}[X]/\langle N(X) \rangle$. We define a map $\kappa:\mathbb{F}_{q^m}^{n}\rightarrow \mathbb{F}_{q^m}[X]/\langle N(X) \rangle$. This map is easily verified to be an isomorphism. We define the polynomial associated with the vector $\bm{u}=(u_{0},u_{1},\ldots,u_{n-1})\in \mathbb{F}_{q^m}^{n}$ as $\bm{u}(X)=\sum_{i=0}^{n-1}u_{i}X^{i} \in \phi$. For vectors $\bm{u},\bm{v}\in \mathbb{F}_{q^m}^{n}$, we define their product as $\bm{u} \cdot \bm{v}:=\bm{u}(X)\bm{v}(X)$ mod $N(X)$ $\in \phi$, i.e., 
  \begin{align*}
      \bm{u} \cdot \bm{v} &= \left(\sum_{i =0}^{n-1}u_{i}X_{i}\right)\bm{v}(X) ~\text{mod} ~ N(X)\\
     & = \sum_{i=0}^{n-1}u_{i}\Big(X^{i}\bm{v}(X) ~\text{mod} ~N(X)\Big).
  \end{align*}
\begin{definition}
  (Ideal Matrix) Let $N(X) \in \mathbb{F}_{q}[X]$ be a polynomial of degree $n$ and vector $\bm{v} \in \mathbb{F}_{q^m}^{n}$. The ideal matrix associated with $\bm{v}$ (with respect to $N(X)$) is the $n \times n$ matrix, denoted by $\mathcal{IM}(\bm{v})$, whose $i$-th row (for $i=0,\ldots,n-1$) corresponds to the coefficients of the polynomial $X^{i}v(X)$ mod $N(X)$.

  \begin{equation}
      \mathcal{IM}(\bm{v}) = 
      \begin{pmatrix}
      \bm{v}  \\
       X \bm{v}~ \text{mod} ~N(X)\\
      \vdots\\
      X^{n-1}\bm{v} ~ \text{mod} ~N(X))
      \end{pmatrix}.
      \nonumber
  \end{equation}
\end{definition}

It is easy to verify that $\bm{u} \cdot \bm{v} = \bm{u}  \cdot \mathcal{IM}(\bm{v})=\bm{v} \cdot \mathcal{IM}(\bm{u})=\bm{v} \cdot \bm{u}$. Next, we will present the definition of ideal codes. It has been proven in \cite{ref6} that if $m$ and $n$ are two different prime numbers and $N(X)$ is irreducible, then a non-zero ideal matrix is always non-singular. To reduce the key size, we only consider ideal codes in systematic form.

\begin{definition}
    (Ideal Codes) Let $N(X) \in \mathbb{F}_{q}[X]$ be a polynomial of degree $n$. A code $\mathcal{C}$ over $\mathbb{F}_{q^m}$ with parameters $[sn,tn]$ is called an $(s,t)$-ideal code if it admits a systematic generator matrix of the block form:
\begin{equation}
   \mathbf{G} =
\begin{pmatrix}
& \mathcal{IM}(\bm{g}_{1,1}) & \cdots & \mathcal{IM}(\bm{g}_{1,s-t}) \\
\mathbf{I}_{tn} & \vdots & \ddots & \vdots \\
& \mathcal{IM}(\bm{g}_{t,1}) & \cdots & \mathcal{IM}(\bm{g}_{t,s-t}) \\
\end{pmatrix},
\nonumber
\end{equation}
where $\bm{g}_{i,j}\in \mathbb{F}_{q^m}^{n}$, for $1 \leq i \leq t$, $1 \leq j \leq s-t$. 

In the special case $t=1$, the code has parameters $[sn,n]$ and its parity-check matrix in systematic form can be expressed as 
\begin{equation}
   \mathbf{H} =
\begin{pmatrix}
& \mathcal{IM}(\bm{h}_{1})^{T}  \\
\mathbf{I}_{(s-1)n} & \vdots & \\
& \mathcal{IM}(\bm{h}_{s-1})^{T}  \\
\end{pmatrix},
\nonumber
\end{equation}
\end{definition}
where $\bm{h}_{i}\in \mathbb{F}_{q^m}^{n}$, for $1 \leq i \leq s-1$.

Gabidulin introduced Gabidulin codes in 1985 \cite{ref25}. These codes can be regarded as the analog of Reed-Solomon codes under the rank metric, where standard polynomials are replaced by $q$-polynomials.
\begin{definition}
    ($q$-Polynomials) A polynomial $f(x)$ over $\mathbb{F}_{q^m}$ is called a $q$-polynomial (or linearized polynomial) if it has the form $f(x) = \sum_{i=0}^{n}f_{i}x^{q^{i}}$ with coefficients $f_{i} \in \mathbb{F}_{q^m}$. The $q$-degree of $f(x)$ is $n$ if $f_{n} \neq 0$.
\end{definition}

\begin{definition}
(Gabidulin Code) Let $k \leq n \leq m$, and $\bm{g} = (g_{1},\ldots, g_{n}) \in \mathbb{F}_{q^m}^{n}$ be a vector of full rank, i.e., $wt_{R}(\bm{g}) = n$. The $[n,k]$ Gabidulin codes $Gab_{n,k}(\bm{g})$ over $\mathbb{F}_{q^m}$ with generator vector $\bm{g}$ is the linear code generated by the Moore matrix $Moore(\bm{g},k-1)$ defined as:
\begin{equation*}
Moore(\bm{g},k-1)=
\left[
\begin{array}{ccc}
  g_{1} & \ldots & g_{n}\\
  g_{1}^{[1]} & \ldots & g_{n}^{[1]}\\
  \vdots & \ddots & \vdots\\
  g_{1}^{[k-1]} & \ldots & g_{n}^{[k-1]}
\end{array}
\right],
\end{equation*}
where $[i]:= q^{i}$ denotes the $i$-th Frobenius power.
\end{definition}

\begin{remark}
    The above Gabidulin code $Gab_{n,k}(\bm{g})$ is an MRD code \cite{ref28}, and has rank error correction capacity $\lfloor \frac{n-k}{2} \rfloor$. There exist efficient decoding algorithms for Gabidulin codes as shown in \cite{ref25,ref26,ref27}.
\end{remark}

\begin{theorem}
\cite{ref29}
    The dual code of the Gabidulin codes $Gab_{n,k}(\bm{g})$ is also a Gabidulin code $Gab_{n,n-k}(\hat{\bm{g}}^{[-n+k+1]})$ for some $\hat{\bm{g}} \in Gab_{n,n-1}(\bm{g})^{\bot}$ with $wt_{R}(\hat{\bm{g}}) = n$.
\end{theorem}

\subsection{Hard problems in coding theory}
\label{hard}
Now we define a commonly used hard problem in coding theory under the rank metric as follows.
\begin{definition}
(Rank Syndrome Decoding (RSD) Problem) Let $q,m,n,k,t$ be positive integers, and $H \in \mathbb{F}_{q^m}^{(n-k)\times n}$ be a matrix over $\mathbb{F}_{q^m}$ with full rank, $\bm{s} \in \mathbb{F}_{q^m}^{n-k}$. The RSD problem $RSD(q,m,n,k,t)$ is to find a vector $\bm{e} \in \mathbb{F}_{q^m}^{n}$ such that $wt_{R}(\bm{e}) = t$ and $\bm{s} = \bm{e}H^{T}$.
\end{definition}

The syndrome decoding (SD) problem is a hard problem in coding theory under the Hamming metric which has been proven to be NP-complete in \cite{ref23}. Scholars \cite{ref24} have proven that if there is an efficient probability algorithm that can solve the RSD problem, then there is an efficient probability algorithm to solve the SD problem. Therefore, the RSD problem is widely regarded as a difficult problem by the research community under the rank metric. There are currently two main approaches for solving the RSD problems, namely combinatorial attacks and algebraic attacks. Table \ref{tab:1} and Table \ref{tab:2} summarize some known results of combinatorial attacks and algebraic attacks, where $\omega$ is the linear algebra constant, and $\omega \approx 2.81$.
\begin{table}[H]
\small
\centering
\caption{Some known combinatorial attacks on the RSD problem.}
\label{tab:1}       
\begin{tabular}{lll}
\hline\noalign{\smallskip}
Attacks  & Complexity  \\
\noalign{\smallskip}\hline\noalign{\smallskip}
\cite{ref15}  & $\mathcal{O}\left(min\{m^{3}t^{3}q^{(t-1)(k+1)},(k+t)^{3}t^{3}q^{(t-1)(m-t)}\}\right) $\\
\cite{ref24}  & $\mathcal{O}\left((n-k)^{3}m^{3}q^{min\{t\lceil \frac{mk}{n} \rceil,(t-1)\lceil \frac{m(k+1)}{n}\rceil\}}\right)$ \\
\cite{ref17}  & $\mathcal{O}\left((n-k)^{3}m^{3}q^{t\lceil \frac{m(k+1)}{n}\rceil-m}\right)$ \\
\noalign{\smallskip}\hline
\end{tabular}
\end{table}

\begin{table}[htbp]
\small
\centering
\caption{Some known algebraic attacks on the RSD problem.}
\label{tab:2}       
\begin{tabular}{lll}
\hline\noalign{\smallskip}
Attacks & Conditions & Complexity  \\
\noalign{\smallskip}\hline\noalign{\smallskip}
\cite{ref18} & $m\binom{n-k-1}{t} \geq \binom{n}{t}-1$ & $\mathcal{O}\left(\left(\frac{\left((m+n)t\right)^{t}}{t!}\right)^{\omega}\right) $\\
\cite{ref19} & $m\binom{n-k-1}{t} \geq \binom{n}{t}-1$ & $\mathcal{O}\left(m\binom{n-p-k-1}{t} \binom{n-p}{t}^{\omega -1}\right)$, where \\
 & \multirow{3}{*}{} & $p = min\Big\{1 \leq i \leq n: m\binom{n-k-1}{t} \geq \binom{n-i}{t}-1\Big\}$\\
\cite{ref18} & $m\binom{n-k-1}{t} < \binom{n}{t}-1$ & $\mathcal{O}\left(\left(\frac{((m+n)t)^{t+1}}{(t+1)!}\right)^{\omega}\right)$ \\
\cite{ref19} &  $m\binom{n-k-1}{t} < \binom{n}{t}-1$  & $\mathcal{O}\left(q^{at}m\binom{n-k-1}{t}\binom{n-a}{t}^{\omega -1}\right)$, where  \\
 &    & $a = min \Big\{1 \leq i \leq n: m\binom{n-k-1}{t} \geq \binom{n-i}{t}-1 \Big\}$ \\
\cite{ref19} & $m\binom{n-k-1}{t} < \binom{n}{t}-1$   & $\mathcal{O}\left(\frac{B_{b}\binom{k+t+1}{t}+C_{b}(mk+1)(t+1)}{B_{b}+C_{b}}A_{b}^{2}\right)$, where  \\
 &    &  $A_{b} = \sum_{j=1}^{b}\binom{n}{t}\binom{mk+1}{j}$, \\ 
&  & $B_{b}=\sum_{j=1}^{b}m\binom{n-k-1}{t}\binom{mk+1}{j}$, \\
 &    & $C_{b}=\sum_{j=1}^{b}\sum_{i=1}^{j}(-1)^{i+1}\binom{n}{t+i}\binom{m+i-1}{i}\binom{mk+1}{j-i}$, \\
 &   & $b = min\{0 < a < t+2: A_{a} - 1 \leq B_{a}+C_{a}\}$. \\
\noalign{\smallskip}\hline
\end{tabular}
\end{table}

 Next, we will introduce another difficult problem in coding theory under rank metric, the blockwise rank decoding problem. Before introducing the blockwise rank decoding problem, we need the knowledge of blockwise errors. 
\begin{definition}
    (Blockwise Error \cite{song2025interleaved}) An error $\bm{e}=(\bm{e}_{1},\bm{e}_{2},\ldots,\bm{e}_{l}) \in \mathbb{F}_{q^m}^{n}$ is called a blockwise error if the supports of all subvectors $\bm{e}_{i}$ are pairwise disjoint, i.e., $Supp(\bm{e}_{i}) \cap Supp(\bm{e}_{j})=\{0\}$ for all $i \neq j$. Here, each $\bm{e}_{i}\in \mathbb{F}_{q^m}^{n_{i}}$ has rank weight $r_{i}$, and $n = \sum_{i=1}^{l}n_{i}$, $r=\sum_{i=1}^{l}r_{i}$.
\end{definition}

\begin{definition}
    (Blockwise Rank Decoding (BRD) Problem \cite{song2025interleaved}) Let $G \in \mathbb{F}_{q^m}^{k \times n}$ be a generator matrix of a random $[n,k]_{q^m}$-linear code $\mathcal{C}$, $\bm{y} \in \mathbb{F}_{q^m}^{n}$. Let $\mathcal{S}_{\bm{\rho}}^{\bm{\eta}}(BW)$ denote the set of all vectors of length $n$ and with rank weight $r$, where $\bm{\eta} = (n_{1},n_{2},\ldots,n_{l}) \in \mathbb{N}^{l}$, $\bm{\rho} = (r_{1},r_{2},\ldots,r_{l})\in \mathbb{N}^{l}$, and $n=\sum_{i=1}^{l}n_{i}$, $r=\sum_{i=1}^{l}r_{i}$. The $BRD(q,m,n,k,r,\bm{\eta},\bm{\rho})$ problem is to find a vector $\bm{x} \in \mathbb{F}_{q^m}^{k}$ and a blockwise error $\bm{e}\in \mathcal{S}_{\bm{\rho}}^{\bm{\eta}}(BW)$ such that $\bm{y}=\bm{x}G+\bm{e}$.
\end{definition}

\begin{definition}
  (Blockwise RSD (BRSD) Problem)  Let $H \in \mathbb{F}_{q^m}^{(n-k)\times n}$ be a parity-check matrix of a random $[n,k]_{q^m}$-linear code $\mathcal{C}$, $\bm{s} \in \mathbb{F}_{q^m}^{n-k}$. The $BRSD(q,m,n,k,r,\bm{\eta},\bm{\rho})$ problem is to find a blockwise error $\bm{e}\in \mathcal{S}_{\bm{\rho}}^{\bm{\eta}}(BW)$ such that $\bm{s}^{T}=H\bm{e}^{T}$.
\end{definition}

\begin{definition}
    (Ideal BRSD (IBRSD) Problem) Let $H$ be the systematic parity-check matrix of a random $[sn,n]_{q^m}$-ideal code, $\bm{y} \in \mathbb{F}_{q^m}^{(s-1)n)}$. The $IBRSD(q,m,n,k,r,\bm{\eta},\bm{\rho})$ problem is to find a blockwise error $\bm{e}\in \mathcal{S}_{\bm{\rho}}^{\bm{\eta}}(BW)$ such that $\bm{y}^{T}=H\bm{e}^{T}$.
\end{definition}

\begin{definition}
    (Decisional IBRSD (DIBRSD) Problem) Let $H$ be the systematic parity-check matrix of a random $[sn,n]_{q^m}$-ideal code. The $DIBRSD(q,m,n,k,r,\bm{\eta},\bm{\rho})$ problem is hard, if for any probabilistic polynomial time adversary $\mathcal{A}$, the following advantage is negligible:
    \begin{align*}
        Adv_{\mathcal{A}}^{s-DIBRSD(q,m,n,k,r,\bm{\eta},\bm{\rho})}(\lambda):&=\bigg| Pr\left[\mathcal{A}(H,\bm{y})=1\Big|H \stackrel{\text{\$}}{\leftarrow} \mathbb{F}_{q^m}^{(s-1)n\times sn}, \bm{y} \stackrel{\text{\$}}{\leftarrow} \mathbb{F}_{q^m}^{(s-1)n}\right]\\
        &-Pr\left[\mathcal{A}(H,H\bm{e}^{T})=1|H \stackrel{\text{\$}}{\leftarrow} \mathbb{F}_{q^m}^{(s-1)n \times sn},\bm{e} \stackrel{\text{\$}}{\leftarrow} \mathcal{S}_{\bm{\rho}}^{\bm{\eta}}(BW)\right] \bigg|
    \end{align*}
\end{definition}

The above hard problem has only one syndrome. Next, we will introduce another hard problem in coding theory, which has multiple syndromes.

\begin{definition}
    (Rank Support Learning (RSL) Problem) Given $(\bm{H},\bm{S}) \in \mathbb{F}_{q^m}^{(n-k)\times n} \times \mathbb{F}_{q^m}^{l \times (n-k)}$, the $RSL(m,n,k,\omega,l)$ problem is to compute a $\mathbb{F}_{q}$-subspace $\mathcal{V} \subset \mathbb{F}_{q^m}$ of dimension $\omega$ such that there exists a matrix $E \in \mathcal{V}^{l \times n}$ satisfying $HE^{T} = S^{T}$.
\end{definition}

Reference \cite{ref7} proposes a new error type, called non-homogeneous errors, which are presented as follows.

\begin{definition}
    (Non-Homogeneous Errors) An error $\bm{e} = (\bm{e}_{1},\bm{e}_{2},\bm{e}_{3}) \in \mathbb{F}_{q^m}^{n}$ is called a non-homogeneous error if \\
    1) its sub-errors $(\bm{e}_{1},\bm{e}_{3})\in \mathbb{F}_{q^m}^{n_{1}}\times \mathbb{F}_{q^m}^{n_{3}}$ have rank weight $r_{1}$, \\
    2) its sub-errors $\bm{e}_{2}\in \mathbb{F}_{q^m}^{n_{2}}$ have rank weight $r_{2}$, \\
    3) the support of the sub-error $\bm{e}_{2}\in \mathbb{F}_{q^m}^{n_{2}}$ contains the support of the sub-error $(\bm{e}_{1},\bm{e}_{3})$,\\
    where the vectors $\bm{e}_{1}\in \mathbb{F}_{q^m}^{n_{1}}$, $\bm{e}_{2}\in \mathbb{F}_{q^m}^{n_{2}}$, $\bm{e}_{3}\in \mathbb{F}_{q^m}^{n_{3}}$, and $n=n_{1}+n_{2}+n_{3}$.
\end{definition}

 Let $\mathcal{S}_{\bm{\rho}}^{\bm{\eta}}(NH)$ denote the set of all such non-homogeneous errors, then we have 
\begin{equation}
\nonumber
 \mathcal{S}_{\bm{\rho}}^{\bm{\eta}}(NH)=\bigg\{\bm{e}=({\bm{e}_{1},\bm{e}_{2},\bm{e}_{3}})\Big|\bm{e}_{1}\in \mathbb{F}_{q^m}^{n_{1}},\bm{e}_{2}\in \mathbb{F}_{q^m}^{n_{2}},\bm{e}_{3}\in \mathbb{F}_{q^m}^{n_{3}},wt_{R}\Big((\bm{e}_{1},\bm{e}_{3})\Big)=r_{1},wt_{R}(\bm{e}_{2})=r_{2},Supp\Big((\bm{e}_{1},\bm{e}_{3})\Big)\subset Supp(e_{2})\bigg\}.
\end{equation}

Similarly, regarding the non-homogeneous error, we have the following hard problems.

 \begin{definition}
     (Non-Homogeneous Rank Decoding (NHRD) Problem) Let $G \in \mathbb{F}_{q^m}^{k \times n}$ be a generator matrix of a random $[n,k]_{q^m}$-linear code $\mathcal{C}$, $\bm{y} \in \mathbb{F}_{q^m}^{n}$. The $NHRD(q,m,n,k,\bm{\eta},\bm{\rho})$ problem is to find a vector $\bm{x} \in \mathbb{F}_{q^m}^{k}$ and a non-homogeneous error $\bm{e}\in \mathcal{S}_{\bm{\rho}}^{\bm{\eta}}(NH)$ such that $\bm{y}=\bm{x}G+\bm{e}$.
 \end{definition}

\begin{definition}
   (Non-Homogeneous RSD (NHRSD) Problem)  Let $H \in \mathbb{F}_{q^m}^{(n-k)\times n}$ be a parity-check matrix of a random $[n,k]_{q^m}$-linear code $\mathcal{C}$, $\bm{s} \in \mathbb{F}_{q^m}^{n-k}$. The $NHRSD(q,m,n,k,\bm{\eta},\bm{\rho})$ problem is to find a non-homogeneous error $\bm{e}\in \mathcal{S}_{\bm{\rho}}^{\bm{\eta}}(NH)$ such that $s^{T}=H\bm{e}^{T}$.
\end{definition}

\begin{definition} (Non-Homogeneous RSL (NHRSL) Problem) Let $H \in \mathbb{F}_{q^m}^{(n-k)\times n}$ and $S \in \mathbb{F}_{q^m}^{N\times (n-k)}$. Suppose that $\bm{\eta}=(n_1,n_2,n_3)$ is a partition of $n$ and $\bm{\rho}=(r_1,r_2)$ with $r_1 \le r_2 \le m$. The Non-Homogeneous Rank Support Learning problem $NHRSL(m,\bm{\eta},k,\bm{\rho},N)$ consists in finding $\mathbb{F}_q$-subspaces $\mathcal{V}_1 \subset \mathcal{V}_2 \subseteq \mathbb{F}_{q^m}$ with $\dim_{\mathbb{F}_q}(\mathcal{V}_1)=r_1$, $\dim_{\mathbb{F}_q}(\mathcal{V}_2)=r_2$, for which there exists a matrix $E = [\,E_1 \mid E_2 \mid E_3\,]$ satisfying $E_1 \in \mathcal{V}_1^{N\times n_1}$, $E_2 \in \mathcal{V}_2^{N\times n_2}$, $E_3 \in \mathcal{V}_1^{N\times n_3}$ and $H E^{\mathsf T} = S^{\mathsf T}$.
\end{definition}

\subsection{Public Key Encryptions and Key Encapsulation Mechanisms}
\subsubsection{Public key encryption}

A public key encryption (PKE) scheme is defined by a tuple of four polynomial-time algorithms PKE=($\textbf{Setup}$, $\textbf{KGen}$, $\textbf{Enc}$, $\textbf{Dec})$, where

$\bullet$ $\textbf{Setup}$, input the security parameters $\lambda$ and output the global parameter $\pi$, i.e., $\pi \gets$ $\textbf{Setup}(1^\lambda)$.

$\bullet$ $\textbf{KGen}$, input the global parameter $\pi$ and output the public key $pk$ and secert key $sk$, i.e., $(pk,sk) \gets$ $\textbf{KGen}(\pi) $.

$\bullet$ $\textbf{Enc}$, input the plaintext $\bm{m}$ and the public key $pk$, output the ciphertext $\bm{c}$, i.e., $\bm{c} \gets$ \textbf{Enc}$(pk,\bm{m}) $.

$\bullet$ $\textbf{Dec}$, input the ciphertext $\bm{c}$ and the secret key $sk$, output the plaintext $\bm{m}$ or $\bot$, i.e., $ \bm{m} \gets $\textbf{Dec}$(\bm{c},sk)$ or $\bot \gets $\textbf{Dec}$(\bm{c},sk)$ .
\subsubsection{Key encapsulation mechanism}
A key encapsulation mechanism (KEM) scheme with a key space $\mathcal{K}$ consists of three polynomial-time algorithms, i.e., KEM=(\textbf{KGen}, \textbf{Encapsulate}, \textbf{Decapsulate}),

$\bullet$ \textbf{KGen}, input the security parameter $\lambda^{'}$ and output the public key $pk^{'}$ and secert key $sk^{'}$, i.e., $(pk^{'},sk^{'}) \gets $\textbf{KGen}$(1^{\lambda^{'}})$.

$\bullet$ \textbf{Encapsulate}, input the public key $pk^{'}$, output an encapsulation $\bm{c}^{'}$ and a key $K\in \mathcal{K}$, i.e., $(\bm{c}^{'},K) \gets $\textbf{Encapsulate}$(pk^{'})$.

$\bullet$ \textbf{Decapsulate}, input the encapsulation $\bm{c}^{'}$ and the secret key $sk^{'}$, output a key $K$, i.e., $K \gets$ \textbf{Decapsulate}$(\bm{c}^{'},sk^{'})$.

\subsection{IND-CPA and IND-CCA2}
\label{cpa}
\subsubsection{IND-CPA}
Indistinguishability under chosen-plaintext attack (IND-CPA) is a security concept for general-use public key encryption (PKE) scheme. We describe IND-CPA as a game in Table \ref{tab:3}. The goal of this game is to break the indistinguishability property of the encryption scheme under a chosen-plaintext attack. Firstly, we need to define two characters, the challenger $C$ and the adversary $\mathcal{I}$. The challenger's task is to act as a referee, host the game, and provide feedback on the adversary's behavior. The adversary attacks the current system with the method of choosing plaintext attacks. The details of the game are showing in Table \ref{tab:3}:

\begin{table}[htbp]
    \centering
    \caption{The IND-CPA GAME.}
    \label{tab:3}
    \renewcommand{\arraystretch}{1.2} 
    \begin{tabular}{|p{0.8cm}|p{7cm}|} 
        \hline
        \textbf{Step 1} & 
        The challenger $C$ creates the IND-CPA system and sends the public key $pk$ to the adversary $\mathcal{I}$. \\
        \hline
        \textbf{Step 2} & 
        The adversary $\mathcal{I}$ can make adaptive decryption queries to the  challenger. \\
        \hline
        \textbf{Step 3} & 
       The adversary $\mathcal{I}$ selects two different plaintexts $\bm{x}_{0}$ and $\bm{x}_{1}$ of same length and sends them to challenger $C$. \\
        \hline
        \textbf{Step 4} & 
        The challenger $C$ randomly selects $b \in \{0,1\}$ and encrypts $\bm{x}_{b}$ as $\bm{c}$=Encryption$(pk,\bm{x}_{b})$, then sends the ciphertext $\bm{c}$ to the adversary $\mathcal{I}$. \\
        \hline
         \textbf{Step 5} & 
        The adversary $\mathcal{I}$ can make adaptive decryption queries to the challenger as the step 2, except that it cannot perform decryption queries on $\bm{c}$. \\
        \hline
        \textbf{Step 6} & 
        The adversary $\mathcal{I}$ guesses the plaintext encrypted by the challenger $C$ in the Step 4 is $\bm{x}_{0}$ or $\bm{x}_{1}$, and outputs the guessed result as $b^{'}$. If $b^{'} = b$, then the adversary $\mathcal{I}$ wins. \\
        \hline
    \end{tabular}
\end{table}

\begin{remark}
To ensure stronger security, Goldwasser and Micali introduced probabilistic encryption and a stronger security notion called semantic security (SS) in 1984, which is theoretically equivalent to IND-CPA. 
\end{remark}
\subsubsection{IND-CCA2}
The indistinguishability under adaptive chosen-ciphertext
attack 2 (IND-CCA2) is the strongest security concept for general-use public key
encryption (PKE) schemes. We describe IND-CCA2 as a game in Table \ref{tab4}. The security goal of IND-CCA2 is to ensure that an adversary cannot distinguish between the ciphertexts of two chosen plaintexts, even when given access to a decryption oracle. This attack model is adaptive, meaning the adversary can make decryption queries interleaved with its challenges, dynamically adjusting its strategy based on previous outputs. Its core logic is defined through a challenger-adversary game, whose specific details are presented in Table \ref{tab4}. Furthermore, there are general conversions \cite{ref30,ref5,ref31,ref32} that can transform an IND-CPA secure PKE into IND-CCA2 secure. 

\begin{table}[htbp]
    \centering
    \caption{The IND-CCA2 GAME.}
    \label{tab4}
    \renewcommand{\arraystretch}{1.5} 
    \begin{tabular}{|m{3.5cm}|p{10cm}|} 
        \hline
        \textbf{Initialization Phase} & 
        The challenger generates the public key ($pk$) and private key ($sk$) of the encryption scheme, sends the public key $pk$ to the adversary, and retains the private key $sk$ for itself. \\
        \hline
        \textbf{Preprocessing Query Phase} & 
       The adversary may select any ciphertexts and submit them to the challenger. The challenger decrypts these ciphertexts using the private key $sk$ and returns the resulting plaintexts to the adversary. (During this phase, the adversary can obtain decryption patterns through multiple queries.)\\
        \hline
        \textbf{Challenge Phase} & 
       The adversary selects two plaintexts ($m_{0}$ and $m_{1}$) of the same length and submits them to the challenger. The challenger randomly selects a bit $b \in \{0, 1\}$, encrypts $m_b$ with the public key $pk$ to generate the challenge ciphertext $c^{*}$, and returns $c^{*}$ to the adversary. (The adversary does not know the value of $b$ and aims to infer it.) \\
        \hline
        \textbf{Post-Processing Query Phase} & 
        The adversary may continue to submit ciphertexts to the challenger for decryption, but must not submit the challenge ciphertext $c^{*}$. The challenger still returns the decryption results. The adversary combines all query information and finally outputs a guess $b^{'}$ for the value of $b$. If $b^{'} = b$, then the adversary $\mathcal{I}$ wins.\\
        \hline
    \end{tabular}
\end{table}

\subsection{RQC Cryptosystem}
We denote the following sets:

\begin{align*}
    \mathcal{S}_{\gamma,1}^{n}(\mathbb{F}_{q^m})&=\{\bm{x}\in \mathbb{F}_{q^m}^{n}\big|dim_{\mathbb{F}_{q}}(Supp(x))=\gamma,1\in Supp(\bm{x})\},\\
    \mathcal{S}_{\gamma}^{n}(\mathbb{F}_{q^m})&=\{\bm{x}\in \mathbb{F}_{q^m}^{n}\big|dim_{\mathbb{F}_{q}}(Supp(x))=\gamma)\}.
\end{align*}

Next, we briefly review the classic RQC cryptosystem proposed by \cite{ref6} as shown in Fig.\ref{fig123}.
\begin{figure}[h]
\begin{minipage}{\linewidth}
\begin{center}
    \fbox{ 
        \begin{minipage}{0.9\textwidth} 
            \underline{Setup($1^{\lambda}$):}\\
            \hspace{2em} $\bullet$ Generates and outputs the global parameters $param = (n,k,\gamma,\omega_{r},N(X))$, where $N(X) \in \mathbb{F}_{q}[X]$ is an irreducible polynomial of degree $n$.\\
            \underline{KGen(param):}\\
            \hspace{1cm} $\bullet$ Sample $\bm{h} \stackrel{\text{\$}}{\leftarrow}\mathbb{F}_{q^m}$, $\bm{g} \stackrel{\text{\$}}{\leftarrow} \mathcal{S}_{n}^{n}(\mathbb{F}_{q^m})$, and $(\bm{x},\bm{y}) \stackrel{\text{\$}}{\leftarrow} \mathcal{S}_{\gamma,1}^{2n}(\mathbb{F}_{q^m})$\\
            \hspace{2em} $\bullet$ Compute the generator matrix $G \in \mathbb{F}_{q^m}^{k \times n}$ of the $[n,k]_{q^m}$-Gabidulin codes $\mathcal{G}_{\bm{g}}$ with the generator vector $\bm{g}$\\
            \hspace{2em} $\bullet$ Compute $\bm{s} =\bm{x}+\bm{h} \cdot \bm{y}$ mod $N(X)$\\
            \hspace{2em} $\bullet$ Output a public key $pk = (\bm{g},\bm{h},\bm{s})$ and a private key $sk = (\bm{x},\bm{y})$\\
            \underline{Enc(pk,$\bm{m}$,$\theta$):}\\
             \hspace{2em} $\bullet$ Input the public key $pk$ and a plaintext $\bm{m} \in \mathbb{F}_{q^m}^{k}$\\
               \hspace{2em} $\bullet$ Use the randomness $\theta$ to generate $(\bm{e},\bm{r}_{1},\bm{r}_{2}) \stackrel{\text{\$}}{\leftarrow} \mathcal{S}_{\omega_{r}}^{3n}(\mathbb{F}_{q^m})$\\
              \hspace{2em} $\bullet$ Compute $\bm{u}=\bm{r}_{1}+\bm{h} \cdotp \bm{r}_{2}$ mod $N(X)$ and $\bm{v} = \bm{m}G+\bm{s} \cdotp \bm{r}_{2}+\bm{e}$ mod $N(X)$  \\
              \hspace{2em} $\bullet$  Output a ciphertext $\bm{c} = (\bm{u},\bm{v})$  \\
            \underline{Dec(sk, $\bm{c}$):}\\
             \hspace{2em} $\bullet$ Input the private key $sk$ and the ciphertext $\bm{c}$\\
             \hspace{2em} $\bullet$ Output the plaintext $\bm{m} := \mathcal{G}_{\bm{g}}.Decode(\bm{v}-\bm{y} \cdotp \bm{u} ~mod ~N(X))$
        \end{minipage} }
\end{center}
\end{minipage}
 \caption{The classic RQC cryptosystem}
    \label{fig123}
\end{figure}

\section{Gabidulin-Kronecker product codes}
\label{sec3333}
In this section, we will present more precise information about the minimum rank distance of the Gabidulin-Kronecker product codes. Before that, we need to provide an upper bound for the rank distance of a rank code, i.e., Singleton-Style Bound.

\begin{lemma} \cite{ref41}
    The minimum rank distance $d$ of any linear $[n,k]_{q^m}$ code $\mathcal{C} \subset \mathbb{F}_{q^m}^{n}$ satisfies the following bound:
    \begin{align}
      \label{bound1}  km &\leq (m-d+1)n, &if ~n>m\\ 
      \label{bound2}  k &\leq n-d+1, &if~ n\leq m
    \end{align}
\end{lemma}

\begin{definition}
\label{def21}
    A code $\mathcal{C}$ is called an MRD code if the above bound (\ref{bound1}), (\ref{bound2}) are reached.
\end{definition}

To analyze the minimum distance of Gabidulin-Kronecker product codes, we recall the concept of reducible rank codes, which serve as a superclass containing our codes of interest.

\begin{definition} \cite{ref41}
    A code $\mathcal{C}$ is called a reducible rank code if its generator matrix $G^{'}$ has the following form:
\begin{equation}
\nonumber
    G^{'} = 
    \left[
\begin{array}{cccccc}
G_{1} & 0 & 0 & \ldots & 0 & 0\\
G_{21} & G_{2} & 0 & \ldots & 0 & 0\\
G_{31} & G_{32} & G_{3} & \ldots & 0 & 0\\
\vdots &  \vdots & \vdots& \ddots & \vdots & \vdots \\
G_{r-1,1} & G_{r-1,2} & G_{r-1,3} & \ldots & G_{r-1} & 0\\
G_{r,1} & G_{r,2} & G_{r,3} & \ldots & G_{r,r-1} & G_{r}\\
\end{array}
\right] 
\end{equation}
where matrices $G_{1},G_{2},\ldots,G_{r}$ are generator matrices of rank metric codes with parameters $[n_{i},k_{i}]$ over the finite field $\mathbb{F}_{q^m}$, for $1 \leq i \leq r$. The matrices $G_{a,b} \in \mathbb{F}_{q^m}^{k_{a}\times n_{b}}$, for $a = 2,\ldots,r$ and $b = 1,\ldots,r-1$.
\end{definition}

\begin{definition}
\label{def12}
    (Gabidulin-Kronecker product code \cite{sun2024}) 
     Let $\mathcal{C}_{1}$ be an $[n_{1},k_{1},d_{1}]$ Gabidulin code generated by the matrix $G_{1} = [g_{ij}] \in \mathbb{F}^{k_{1}\times n_{1}}_{q^m}$, and $\mathcal{C}_{2}$ be an $[n_{2},k_{2},d_{2}]$ Gabidulin code with error correcting capability up to $t_{2} = \lfloor \frac{n_{2}-k_{2}}{2} \rfloor$ errors generated by the matrix $G_{2} \in \mathbb{F}^{k_{2}\times n_{2}}_{q^m}$. Then an $[n,k]$ Gabidulin-Kronecker product code $\mathcal{C} = \mathcal{C}_{1} \otimes \mathcal{C}_{2}$ is defined as the linear code with generator matrix 
\begin{equation}
    G = G_{1} \otimes G_{2} =
\left[
\begin{array}{cccc}
g_{11}G_{2} & \ldots & g_{1n_{1}}G_{2}\\
\vdots &  \ddots & \vdots \\
g_{k_{1}1}G_{2} & \ldots & g_{k_{1}n_{1}}G_{2}
\end{array}
\right] \in \mathbb{F}_{q^m}^{k \times n},
\nonumber 
\end{equation}
\end{definition}
\noindent where $n = n_{1}n_{2}$, $k = k_{1}k_{2}$, and $1 \leq i \leq k_{1}$, $1 \leq j \leq n_{1}$. 

We rewrite matrix $G$ as:
 \begin{align*}
    G &= G_{1} \otimes G_{2} \\
    &=
\left[
\begin{array}{cccc}
g_{11}G_{2} & \ldots & g_{1n_{1}}G_{2}\\
\vdots &  \ddots & \vdots \\
g_{k_{1}1}G_{2} & \ldots & g_{k_{1}n_{1}}G_{2}
\end{array}
\right]  \\
&=
\left[
\begin{array}{cccc}
g_{11}I_{k_{2}} & \ldots & g_{1n_{1}}I_{k_{2}}\\
\vdots &  \ddots & \vdots \\
g_{k_{1}1}I_{k_{2}} & \ldots & g_{k_{1}n_{1}}I_{k_{2}}
\end{array}
\right] 
\left[
\begin{array}{cccc}
G_{2} & \ldots & \bm{0}\\
\vdots &  \ddots & \vdots \\
\bm{0} & \ldots & G_{2}
\end{array}
\right] \\
& \triangleq {G}^{'}_{1}{G}^{'}_{2}.
\nonumber 
\end{align*}

As can be shown in Definition \ref{def21}, MRD codes can be obtained in two cases. When $n>m$, the equation $km=(m-d+1)n$ is satisfied. And when $n<m$, the equation $k=n-d+1$ is satisfied. Next, we will also discuss it in two cases. In the first case, we consider $n>m$, with the other parameters matching those in Corollary 1 of Reference \cite{ref41}. In the second case, we consider $n\leq m$, with the other parameters satisfy the conditions in Definition \ref{def12}. Next, we will provide detailed information about the two cases.

\textbf{Case 1}: Let $n_{1} = k_{1},n_{2} = m < n_{1}n_{2}$.

\begin{lemma}
    An $[n,k]$ Gabidulin-Kronecker product code $\mathcal{C} = \mathcal{C}_{1} \otimes \mathcal{C}_{2} $ defined in Definition \ref{def12} is a subcode of a reducible rank code with $G_{i,j} = \bm{0}$, $i>j$, and $G_{2} = G_{1} = G_{3} = \cdots = G_{r}$.
\end{lemma}
\begin{proof}
 Let $G_{i,j} = \bm{0}$, $i>j$, $k_{1}=n_{1}=r$, and $G_{2} = G_{1} = G_{3} = \cdots = G_{r}$. Therefore the generator matrix $G^{'}$ of the reducible rank code $\mathcal{C}^{'}$ has the following form:
\begin{equation}
\nonumber
    G^{'} = 
    \left[
\begin{array}{cccccc}
G_{2} & 0 & 0 & \ldots & 0 & 0\\
0 & G_{2} & 0 & \ldots & 0 & 0\\
0 & 0 & G_{2} & \ldots & 0 & 0\\
\vdots &  \vdots & \vdots& \ddots & \vdots & \vdots \\
0 & 0 & 0 & \ldots & G_{2} & 0\\
0 & 0 & 0 & \ldots & 0& G_{2}\\
\end{array}
\right] .
\end{equation}
  Let $\bm{c}$ be a codeword of Gabidulin-Kronecker product code $\mathcal{C}$, then there exists $\bm{x} \in \mathbb{F}_{q^m}^{k}$ such that $\bm{c} = \bm{x}G=\bm{x}G^{'}_{1}G^{'}_{2}$. Let $\bm{s} = \bm{x}G^{'}_{1}$, therefore we have $\bm{c} = \bm{s}G^{'}_{2} \in \mathcal{C}^{'}$, this means an $[n,k]$ Gabidulin-Kronecker product code $\mathcal{C} = \mathcal{C}_{1} \otimes \mathcal{C}_{2} $ is a subcode of a reducible rank code.
\end{proof}

References \cite{ref41} state that the reducible rank codes are MRD codes under certain conditions, and we will prove below that Gabidulin-Kronecker product codes given in Definition \ref{def12} are also MRD codes when certain conditions are satisfied.

\begin{theorem}
\label{the1}
 Let $\mathcal{C}_{1}$ be an $[n_{1}=k_{1},k_{1},d_{1}]$ Gabidulin code, and $\mathcal{C}_{2}$ be an $[n_{2}=m,k_{2},d_{2}]$ Gabidulin code. Let $\mathcal{C} = \mathcal{C}_{1}\otimes \mathcal{C}_{2}$ be the $[n=n_{1}n_{2},k=k_{1}k_{2},d]$ Gabidulin-Kronecker product code. The minimum rank distance $d$ of $\mathcal{C}$ is $d_{2}$.    
\end{theorem}
  \begin{proof}
 We first prove the lower bound. Let $\bm{c} \in \mathcal{C}$ be a nonzero codeword, therefore there exists one nonzero vector $\bm{x} = (\bm{x}_{1},\bm{x}_{2},\ldots,\bm{x}_{k_{1}}) \in \mathbb{F}_{q^m}^{k_{1}k_{2}}$ such that $\bm{c} = \bm{x}G = (\bm{c}_{1},\ldots,\bm{c}_{n_{1}})$, where $\bm{x}_{i} \in \mathbb{F}_{q^m}^{k_{2}}$, for $1 \leq i \leq k_{1}$, $\bm{c}_{j} \in \mathbb{F}_{q^m}^{n_{2}}$, for $1 \leq j \leq n_{1}$. And we have 
 \begin{align*}
\begin{cases}
     \bm{c}_{1} = (\bm{x}_{1}g_{11}+\ldots+\bm{x}_{k_{1}}g_{k_{1}1})G_{2}\\
     \bm{c}_{2} = (\bm{x}_{1}g_{12}+\ldots+\bm{x}_{k_{1}}g_{k_{1}2})G_{2}\\
     ~~~~\vdots\\
     \bm{c}_{n_{1}} = (\bm{x}_{1}g_{1n_{1}}+\ldots+\bm{x}_{k_{1}}g_{k_{1}n_{1}})G_{2}
\end{cases}
 \end{align*}
Since $\bm{c} \neq \bm{0}$, there exists at least one index $j$ for which the linear combination $\sum_{i=1}^{k_{1}}\bm{x}_{i}g_{ij} \neq 0$. For this $j$, we have $\bm{c}_{j} = (\sum_{i=1}^{k_{1}}\bm{x}_{i}g_{ij})G_{2}$, which is a non-zero codeword of $\mathcal{C}_{2}$. Consequently, $wt_{R}(\bm{c})\ge wt_{R}(\bm{c}_{i}) \ge d_{2}$, establishing the lower bound $d \ge d_{2}$.

 The upper bound is proved as follows. Since $\mathcal{C}_{1}$ is a Gabidulin code, therefore we can obtain $d_{1}=n_{1}-k_{1}+1 = 1$. The upper bound follows directly from the general fact that the minimum distance of a Kronecker product code cannot exceed the product of the minimum distances of the component codes \cite{sun2024}, i.e., $d \leq d_{1}d_{2}=d_{2}$.

 Overall, we ultimately obtained $d=d_{2}$.   $\hfill\blacksquare$
  \end{proof}
\begin{remark}
 The aforementioned Theorem \ref{the1} implies that the Gabidulin-Kronecker product codes are MRD codes when $n_{1}=k_{1},n_{2}=m < n_{1}n_{2}$ holds.
\end{remark}

\textbf{Case 2}: Let $k_{1} \leq n_{1},k_{2} \leq n_{2}, n_{1}n_{2} \leq m$.

Without loss of generality, we assume $d_{1}\leq d_{2}$, otherwise we will use $\mathcal{C}_{2} \otimes \mathcal{C}_{1}$ in this paper. The following Theorem \ref{pro1} gives the bound for the  minimum rank distance of the Gabidulin-Kronecker product code.

\begin{theorem} \cite{sun2024}
 \label{pro1}
    Let $\mathcal{C}_{1}$ be an $[n_{1},k_{1},d_{1}]$ Gabidulin code, and $\mathcal{C}_{2}$ be an $[n_{2},k_{2},d_{2}]$ Gabidulin code. Let $\mathcal{C}$ be the $[n,k,d]$ Gabidulin-Kronecker product code given in Definition \ref{def12}. The minimum rank distance $d$ of $\mathcal{C}$ satisfies $d_{2} \leq d \leq d_{1}d_{2}$.
\end{theorem}
\begin{remark}
   The bounds in Theorem \ref{pro1} are known to be tight \cite{sun2024}. We have further verified this tightness via extensive Magma experiments (over 1000 tests), which confirmed the existence of GK product codes whose minimum distance achieves both the lower bound $d_{2}$ and the upper bound $d_{1}d_{2}$.
\end{remark}

\section{Extended Gabidulin-Kronecker product codes}
\label{sec444}
In this section, we will introduce a new rank-metric code. Before that, we first present some information about Extended Gabidulin codes.

\begin{definition}
  (Extended Gabidulin (EG) codes\cite{berger2009construction})    Let $q,m,n,t,k$ be integers and $k \leq t \leq min\{n,m\}$. Let $\bm{g} = (g_{1},g_{2},\ldots,g_{n}) \in \mathbb{F}_{q^m}^{n}$ be the generator with weight $t$. An EG code of dimension $k$ and length $n$ generated by $\bm{g}$ is defined as $$EG_{k}({\bm{g}}) = \Big\{f(\bm{g})= \Big(f(g_{1}),f(g_{2}),\ldots,f(g _{n})\Big):f(x) \in \mathcal{L}_{\leq k-1}[x]\Big\}.$$ 
\end{definition}

The following properties of EG codes are known \cite{song2025interleaved}.

\begin{proposition} \cite{song2025interleaved}
    The dimension of the $EG_{k}(\bm{g})$ codes is $k$.
\end{proposition}

\begin{proposition} \cite{song2025interleaved}
    The minimal distance of the $EG_{k}(\bm{g})$ codes is $t-k+1$.
\end{proposition}

Next, we will introduce the new rank metric code, namely Extended Gabidulin-Kronecker product codes.
\begin{definition}
  (Extended Gabidulin-Kronecker product codes)  
\label{EGK}
Let $q,m,n,k,t_{1},t_{2},n_{1},n_{2},k_{1},k_{2},d_{1},d_{2}$ be integers and $k_{1} \leq t_{1} \leq min \{n_{1},m\}$, $k_{2}\leq t_{2} \leq min\{n_{2},m\}$. Let $\mathcal{C}_{1}$ be an $[n_{1},k_{1},d_{1}]$ Extended Gabidulin code generated by the matrix $G_{1} = [g_{ij}] \in \mathbb{F}_{q^m}^{k_{1}\times n_{1}}$ of weight $t_{1}$, and $\mathcal{C}_{2}$ be an $[n_{2},k_{2},d_{2}]$ Extended Gabidulin code generated by the matrix $G_{2} \in \mathbb{F}_{q^m}^{k_{2}\times n_{2}}$ with generator $\bm{g}^{'}=(g^{'}_{1},g^{'}_{2},\ldots,g^{'}_{n})$ of weight $t_{2}$. The $[n,k]$ Extended Gabidulin-Kronecker (EGK) product code $\mathcal{C} = \mathcal{C}_{1} \otimes \mathcal{C}_{2}$ is defined as the linear code with generator matrix
\begin{equation}
    G = G_{1} \otimes G_{2} =
\left[
\begin{array}{cccc}
g_{11}G_{2} & \ldots & g_{1n_{1}}G_{2}\\
\vdots &  \ddots & \vdots \\
g_{k_{1}1}G_{2} & \ldots & g_{k_{1}n_{1}}G_{2}
\end{array}
\right] \in \mathbb{F}_{q^m}^{k \times n},
\nonumber 
\end{equation}
\noindent where $n = n_{1}n_{2}$ and $k = k_{1}k_{2}$.
\end{definition}

The dimension and minimal distance of the Extended Gabidulin-Kronecker product code $\mathcal{C}$ are presented formally in Proposition \ref{dimension} and Proposition \ref{minimal distance}.
\begin{proposition}\label{dimension}
   The dimension of the Extended Gabidulin-Kronecker product code $\mathcal{C}$ defined in Definition \ref{EGK} is $k=k_{1}k_{2}$. 
\end{proposition}
\begin{proof}
Assuming $G_1$ has row vectors $\bm{g}_1, \bm{g}_2, \ldots, \bm{g}_{k_1}$, where $\bm{g}_{i} \in \mathbb{F}_{q^m}^{n_{1}}$ for $1 \leq i \leq k_{1}$, and $G_2$ has row vectors $\bm{g}_{1}^{'},\bm{g}_{2}^{'},\ldots, \bm{g}_{k_2}^{'}$, where $\bm{g}_{i}^{'} \in \mathbb{F}_{q^m}^{n_{2}}$ for $1 \leq i \leq k_{2}$. According to the definition of the kronecker product, $G=G_{1}\otimes G_{2}$ has $k_{1}k_{2}$ row vectors in total. Our goal is to prove that these $k_{1}k_{2}$ row vectors are linearly independent, thus obtaining the dimension of the kronecker product code $\mathcal{C}$ as $k_{1}k_{2}=k$.    

Assume, for contradiction, that the $k_{1}k_{2}$ rows of $G$ are linearly dependent. Then there exists coefficients $\lambda_{i,j}\in \mathbb{F}_{q^m}$ that is not entirely zero, such that $\sum_{i=1}^{k_{1}}\sum_{j=1}^{k_{2}}\lambda_{i,j}(\bm{g}_{i}\otimes \bm{g}_{j}^{'})=\bm{0}$. Let $\bm{c}_{i} \triangleq \sum_{j=1}^{k_{2}}\lambda_{i,j}\bm{g}_{j}^{'} \in \mathbb{F}_{q^m}^{n_{2}}$ for $1 \leq i \leq k_{1}$. Then the dependency relation becomes 
\begin{equation} \label{equation 1}
   \sum_{i=1}^{k_{1}}\sum_{j=1}^{k_{2}}\lambda_{i,j}(\bm{g}_{i}\otimes \bm{g}_{j}^{'})=\sum_{i=1}^{k_{1}}\bm{g}_{i}\otimes(\sum_{j=1}^{k_{2}}\lambda_{i,j}\
\bm{g}_{j}^{'})= \sum_{i=1}^{k_{1}}\bm{g}_{i}\otimes \bm{c}_{i}=\bm{0}. 
\end{equation}
We claim $\bm{c}_{i}=0$, for $1 \leq i \leq k_{1}$. Otherwise, assuming that there exists one $\bm{c}_{t} \neq \bm{0}$, for $1 \leq t \leq k_{1}$. Since $\bm{g}_{1},\bm{g}_{2},\ldots,\bm{g}_{k_{1}}$ are linearly independent, there exists $p \in \{1,2,\ldots,n_{1}\}$, such that $\bm{g}_{t}(p)\neq 0$ (i.e., the $p$-th element of $\bm{g}_{t}$ is nonzero). We consider the product of the $p$-th element of $\bm{g}_{i}$ and the $q$-th element of $\bm{c}_{i}$, denoted as $\bm{g}_{i}(p) \cdot \bm{c}_{i}(q)$, and obtain $\sum_{i=1}^{k_{1}}\bm{g}_{i}(p)\cdot \bm{c}_{i}(q)=0$ from equation (\ref{equation 1}). Due to $\bm{g}_{t}(p)\neq 0$ and $\bm{c}_{t}\neq \bm{0}$, there exists $q$, such that $\bm{c}_{t}(q) \neq 0$. This means that there exist numbers are not all zeros, which makes $\bm{g}_{1}(p),\bm{g}_{2}(p),\ldots,\bm{g}_{k_{1}}(p)$ are linearly dependent and contradictory to $\bm{g}_{1},\bm{g}_{2},\ldots,\bm{g}_{k_{1}}$ being linearly independent. So we can obtain $\bm{c}_{i} = \bm{0}$, for $1 \leq i \leq k_{1}$.

This means $\sum_{j=1}^{k_{2}}\lambda_{i,j}\bm{g}_{j}^{'} = \bm{0}$, for $1 \leq i \leq k_{1}$. However, $\bm{g}_{1}^{'},\bm{g}_{2}^{'},\ldots, \bm{g}_{k_2}^{'}$ are linearly independent, so $\lambda_{i,j}=0$ for $1 \leq i \leq k_{1}$, $1 \leq j \leq k_{2}$, this is a contradiction. Therefore, the $k_{1}k_{2}$ rows of the kronecker product code $\mathcal{C}$ are linearly independent, and the dimension of code $\mathcal{C}$ is $k$.  $\hfill\blacksquare$
\end{proof}

\begin{proposition}{\label{1}}
Let $A \in \mathbb{F}_{q^m}^{m \times n}$, $B \in \mathbb{F}_{q^m}^{p \times q}$, $C \in \mathbb{F}_{q^m}^{n \times k}$, $D \in \mathbb{F}_{q^m}^{q \times r}$. Then $$(A \otimes B)(C \otimes D)=(AC) \otimes (BD).$$
\end{proposition}
\begin{proof}
    Let $A = [A_{ij}]_{1 \leq i \leq m, 1 \leq j \leq n}$ and $C=[C_{ls}]_{1 \leq l \leq n, 1 \leq s \leq k}$, then  
\begin{equation}
     A \otimes B =
\left[
\begin{array}{cccc}
A_{11}B & \ldots & A_{1n}B\\
\vdots &  \ddots & \vdots \\
A_{m1}B & \ldots & A_{mn}B
\end{array}
\right] \in \mathbb{F}_{q^m}^{mp \times nq},
\nonumber 
C \otimes D =
\left[
\begin{array}{cccc}
C_{11}D & \ldots & C_{1k}D\\
\vdots &  \ddots & \vdots \\
C_{n1}D & \ldots & C_{nk}D
\end{array}
\right] \in \mathbb{F}_{q^m}^{nq \times kr}.
\end{equation} 
\end{proof}
From the above, we obtain $(A \otimes B)( C \otimes D)\in \mathbb{F}_{q^m}^{mp \times kr}$, and the $(i,j)$ block of $(A \otimes B)( C \otimes D)$ is $\sum_{k=1}^{n}(A_{ik}B)(C_{kj}D) = \sum_{k=1}^{n}(A_{ik}C_{kj})BD=(\sum_{k=1}^{n}A_{ik}C_{kj})(BD)$. We denote the $(i,j)$-th element $\sum_{k=1}^{n}A_{ik}C_{kj}$ of matrix $(AC)$ as $(AC)_{ij}$.

According to the definition of kronecker product, we have 
\begin{equation}
    (AC) \otimes (BD) =
\left[
\begin{array}{cccc}
(AC)_{11}(BD) & \ldots & (AC)_{1r}(BD))\\
\vdots &  \ddots & \vdots \\
(AC)_{m1}(BD) & \ldots & AC_{mk}(BD)
\end{array}
\right] \in \mathbb{F}_{q^m}^{mp \times kr}.
\nonumber 
\end{equation}
The $(i,j)$ block of $(AC) \otimes (BD)$ is $(AC)_{ij}\cdotp (BD)$, which is equal to the $(i,j)$ block of $(A \otimes B)( C \otimes D)$. The conclusion follows from the same dimensions and entities on both sides. $\hfill\blacksquare$

\begin{proposition}
    If the matrices $A \in \mathbb{F}_{q^m}^{m \times m}$ and $B \in \mathbb{F}_{q^m}^{n \times n}$ are nonsingular, then $A\otimes B$ is also nonsingular with $(A \otimes B)^{-1} = A^{-1}\otimes B^{-1}$.
\end{proposition}

\begin{proof}
 Based on {Proposition \ref{1}}, we have the following result.   
$$(A\otimes B)(A^{-1}\otimes B^{-1})=(AA^{-1})\otimes (BB^{-1}) = I_{m} \otimes I_{n} = I_{mn},$$
$$(A^{-1}\otimes B^{-1})(A\otimes B)=(A^{-1}A)\otimes (B^{-1}B) = I_{m} \otimes I_{n} = I_{mn}.$$
This implies that $A^{-1}\otimes B^{-1}$ is the unique inverse of $A\otimes B$ under kronecker product. Therefore, $A\otimes B$ is nonsingular. $\hfill\blacksquare$
\end{proof}

\begin{proposition} \label{minimal distance}
 The minimal rank distance of the Extended Gabidulin-Kronecker product code $\mathcal{C}$ defined in Definition \ref{EGK} is $t_{2}-k_{2}+1$ for the case of $k_{1}=t_{1},t_{2}=m<t_{1}t_{2}$. And the  minimal distance $d$ of the Extended Gabidulin-Kronecker product code $d$ satisfies $t_{2}-k_{2}+1 \leq d \leq (t_{1}-k_{1}+1)(t_{2}-k_{2}+1)$ for the case of $k_{1} \leq t_{1},k_{2} \leq t_{2}, t_{1}t_{2} \leq m$.
\end{proposition}

\begin{proof}
 Let $G_{1}=[g_{i,j}]_{1 \leq i \leq k_{1}, 1 \leq j \leq n_{1}}$ (resp. $G_{2}$ ) be the generator matrix of $[n_{1},k_{1},d_{1}]$ (resp. $[n_{2},k_{2},d_{2}]$) Extended Gabidulin codes $\mathcal{C}_{1}$ (resp. $\mathcal{C}_{2}$) with generator $\bm{g}_{1} \in \mathbb{F}_{q^m}^{n_{1}}$ (resp. $\bm{g}_{2} \in \mathbb{F}_{q^m}^{n_{2}}$) of weight $t_{1}$ (resp. $t_{2}$). Assume that the first $t_{1}$ (resp. $t_{2}$) coordinates of $\bm{g}_{1}$ (resp. $\bm{g}_{2}$) are linearly independent. Otherwise, perform an invertible matrix on $G_{1}$ and $G_{2}$, respectively. Let $\bm{\epsilon}_{1} \in \mathbb{F}_{q^m}^{t_{1}}$ (resp. $\bm{\epsilon}_{2} \in \mathbb{F}_{q^m}^{t_{1}}$) be a basis of $supp(\bm{g}_{1})$ (resp. $supp(\bm{g}_{2})$). Let $CM(\bm{g}_{1}) \in \mathbb{F}_{q}^{t_{1} \times n_{1}}$ (resp. $CM(\bm{g}_{2}) \in \mathbb{F}_{q}^{t_{2} \times n_{2}}$) of rank $t_{1}$ (resp. $t_{2}$) be the coefficient matrix of $\bm{g}_{1}$ (resp. $\bm{g}_{2}$) under $\bm{\epsilon}_{1}$ (resp. $\bm{\epsilon}_{2}$) such that $\bm{g}_{1} = \bm{\epsilon}_{1}CM(g_{1})$ (resp. $\bm{g}_{2} = \bm{\epsilon}_{2}CM(\bm{g}_{2})$). There exists an invertible matrix $P_{1} \in \mathbb{F}_{q}^{n_{1} \times n_{1}}$ (resp. $P_{2} \in \mathbb{F}_{q}^{n_{2} \times n_{2}}$) such that the last $n_{1}-t_{1}$ (resp. $n_{2}-t_{2}$) columns of $CM(\bm{g}_{1})$ (resp. $CM(\bm{g}_{2})$) are zeros, i.e., 
\begin{align}
\nonumber
    CM(\bm{g}_{1})P_{1} = [V_{1} | \bm{0}_{t_{1} \times (n_{1}-t_{1})}], \hspace{1cm} V_{1} \in \mathbb{F}_{q}^{t_{1} \times t_{1}}\\
    CM(\bm{g}_{2})P_{2} = [V_{2} | \bm{0}_{t_{2} \times (n_{2}-t_{2})}], \hspace{1cm} V_{2} \in \mathbb{F}_{q}^{t_{2} \times t_{2}}
    \nonumber
\end{align}
Further there exists a vector $\bm{g}_{1}^{'} = (g_{11}^{'},g_{12}^{'},\ldots,g_{1t_{1}}^{'},0,\ldots,0) \in \mathbb{F}_{q^m}^{n_{1}}$ (resp. $\bm{g}_{2}^{'} = (g_{21}^{'},g_{22}^{'},\ldots,g_{2t_{2}}^{'},0,\ldots,0) \in \mathbb{F}_{q^m}^{n_{2}}$) such that $\bm{g}_{1}^{'}=\bm{\epsilon}_{1} CM(\bm{g}_{1})P_{1} = \bm{g}_{1}P_{1}$ (resp. $\bm{g}_{2}^{'}=\bm{\epsilon}_{2} CM(\bm{g}_{2})P_{2} = \bm{g}_{2}P_{2}$), and the vector $(g_{11}^{'},g_{12}^{'},\ldots,g_{1t_{1}}^{'}) = \bm{\epsilon}_{1} V_{1}$ (resp. $(g_{21}^{'},g_{22}^{'},\ldots,g_{2t_{2}}^{'}) = \bm{\epsilon}_{2} V_{2}$).  

Let the matrices $G= G_{1} \otimes G_{2}$, $P= P_{1} \otimes P_{2}$, therefore we have $GP= (G_{1}\otimes G_{2})\cdot (P_{1} \otimes P_{2}) =(G_{1}P_{1})\otimes (G_{2}P_{2})= Moore(\bm{g}_{1}P_{1},k_{1}-1) \otimes Moore(\bm{g}_{2}P_{2},k_{2}-1) = Moore(\bm{g}^{'}_{1},k_{1}-1) \otimes Moore(\bm{g}^{'}_{2},k_{2}-1)$. For rank metric codes over $\mathbb{F}_{q^m}$, the invertible matrix $P$ over the basis field is an isometry. Then the codes defined by $G$ are equivalent to the codes defined by $GP$. If we fully expand the matrix $GP$, we will find that $GP$ has the form of $[V_{1}^{'}|\bm{0}|V_{2}^{'}|\bm{0}|\cdots|V_{t_{1}}^{'}|\bm{0}|\bm{0}|\cdots|\bm{0}]$, where $V_{i}^{'}\in \mathbb{F}_{q^m}^{k_{1}k_{2}\times t_{2}}$, for $1 \leq i\leq t_{1}$. Since the rank weight of $[V_{1}^{'}|\bm{0}|V_{2}^{'}|\bm{0}|\cdots|V_{t_{1}}^{'}|\bm{0}|\bm{0}|\cdots|\bm{0}]$ is equal to the rank weight of $[V_{1}^{'}|V_{2}^{'}|\cdots|V_{t_{1}}^{'}]$, and the matrix $[V_{1}^{'}|V_{2}^{'}|\cdots|V_{t_{1}}^{'}]$ exactly forms a Gabidulin-Kronecker product code of dimension $k_{1}k_{2}$ and length $t_{1}t_{2}$. Known from Section \ref{sec3333}, the Gabidulin-Kronecker product codes have the minimal rank distance $t_{2}-k_{2}+1$ for the case of $k_{1}=t_{1},t_{2}=m<t_{1}t_{2}$. And the  minimal rank distance of the Gabidulin-Kronecker product code $d$ satisfies $t_{2}-k_{2}+1 \leq d \leq (t_{1}-k_{1}+1)(t_{2}-k_{2}+1)$ for the case of $k_{1} \leq t_{1},k_{2} \leq t_{2}, t_{1}t_{2} \leq m$. Therefore, by applying Theorem \ref{the1} and Theorem \ref{pro1} to this equivalent code, we obtain the stated bounds on the minimal rank distance of the original EGK code. Thus we can obtain that the Extended Gabidulin-Kronecker product codes have the minimal rank distance $t_{2}-k_{2}+1$ for the case of $k_{1}=t_{1},t_{2}=m<t_{1}t_{2}$. And the minimal rank distance of the Extended Gabidulin-Kronecker product code $d$ satisfies $t_{2}-k_{2}+1 \leq d \leq (t_{1}-k_{1}+1)(t_{2}-k_{2}+1)$ for the case of $k_{1} \leq t_{1},k_{2} \leq t_{2}, t_{1}t_{2} \leq m$. $\hfill\blacksquare$
\end{proof}

\section{Decoding Extended Gabidulin-Kronecker product codes}
\label{sec555}
\subsection{Decoding Extended Gabidulin codes} \label{sec3}
Reference \cite{song2025interleaved} provides two methods for decoding EG codes, and in this section, we will present a different method for decoding EG codes.

\begin{definition}
    Given $\bm{g}=(g_{1},\ldots,g_{n})\in \mathbb{F}_{q^m}^{n}$ with $wt_{R}(\bm{g}) = t$ (where $k \leq t \leq min\{m,n\}$) and a received word $\bm{y} = (y_{1},\ldots,y_{n})\in \mathbb{F}_{q^m}^{n}$, the decoding EG code problem $DecEGCode(\bm{g},\bm{y})$ asks to find a linearized polynomial $f(x) \in \mathcal{L}_{\leq k-1}$ and an error vector $ \bm{e} \neq \bm{0} \in \mathbb{F}_{q^m}^{n}$ with $wt_{R}(\bm{e}) \leq r$ such that $\bm{y} = f(\bm{g})+\bm{e}$.
\end{definition}

Next, let $\bm{g}$ be the generator of the Extended Gabidulin codes with weight $t$, and let $\bm{e}$ be the error vector of rank weight $r$. There are some cases:

Case 1: $\bm{g} = (g_{1},g_{2},\ldots,g_{t},0,\ldots,0)$, $g_{i} \neq 0$, $1 \leq i \leq t$;
 
Case 2: $\bm{g} = (g_{1},\ldots,g_{t},\ldots,g_{n})$, and $wt_{R}(g_{1},\ldots,g_{t})<t$;

Case 3: $\bm{e} = (e_{1},\ldots,e_{t},0\ldots,0)$, $e_{i} \neq 0$, $1 \leq i \leq t$;

Case 4: $\bm{e} = (e_{1},\ldots,e_{t},\ldots,e_{n})$, and $wt_{R}(e_{1},\ldots,e_{t})<r$.

The decoding strategy depends on the structure of $\bm{g}$ and the support of 
$\bm{e}$. The key observation is that for any $\bm{g}$ of weight $t$, there exists an invertible matrix $P\in \mathbb{F}_{q}^{n \times n}$ such that $\bm{g}P= (g_{1}^{'},\ldots,g_{t}^{'},0,\ldots,0)$ with $\{g_{1}^{'},\ldots,g_{t}^{'}\}$ linearly independent (Case 1). Decoding $\bm{y}$ for original $\bm{g}$ is equivalent to decoding $\bm{y}P$ for the transformed $\bm{g}P$. Therefore, we focus on decoding Case 1. For the error vector, if its support is not contained within the first $t$ coordinates, the components beyond position $t$ reveal part of the error directly, allowing us to puncture the code and reduce the problem to decoding a Gabidulin code of length $t$. Thus, the core challenge is decoding Case 1 with an error Case 3 (error support within the first t positions).

Below we only provide decoding methods for Case 1 and Case 3.

 \begin{definition}
     (Symbolic Product)
     The symbolic product $\otimes$ of two linearized polynomials $f(x)$ and $g(x)$ is defined as $f(x)  \otimes g(x) = f(g(x))$.
 \end{definition}

\begin{definition}
    (Right Symbolic Divisor)
  We call $b(x)$ a right symbolic divisor of $a(x)$, if $a(x) = q(x) \otimes b(x)$ for some $q(x)$, and denote by $q(x), r(x) \leftarrow RDiv(a(x),b(x))$, where  $r(x)$ with $deg_{q}r(x) < deg_{q} a(x)$ denotes a possible remainder.  
\end{definition}

\begin{definition}
    (Left Symbolic Divisor)
  We call $b(x)$ a left symbolic divisor of $a(x)$, if $a(x) = b(x) \otimes q(x)$ for some $q(x)$, and denote by $q(x), r(x) \leftarrow LDiv(a(x),b(x))$, where  $r(x)$ with $deg_{q}r(x) < deg_{q}a(x)$ denotes a possible remainder.  
\end{definition}

For linearized polynomials $a(x)$ and $b(x)$, we call $r(x)$ as the right symbolic greatest common divisor if there exists polynomials 
$u(x)$ and $v(x)$, such that $r(x) = v(x)\otimes a(x) + u(x)\otimes b(x)$. Next, we will give the linearized Extended Euclidean algorithm (LEEA) with a stopping condition $d_{S} > 0$ which can be shown in Algorithm \ref{alg:alg1}.

\begin{algorithm}[h]
\caption{LEEA (Linearized Extended Euclidean Algorithm) with Stopping Condition}
\label{alg:alg1}
\begin{algorithmic}[1]
\State \textbf{Input}: Linearized polynomials \( a(x), b(x) \) with \( \deg_{q} a(x) \geq \deg_{q} b(x) \), and a stopping degree \( d_{stop} > 0 \);
\State \textbf{Initialize}: \( r_{-1}(x) \leftarrow a(x) \), \( r_0(x) \leftarrow b(x) \), \( i \leftarrow 1 \), \( u_{-1}(x) \leftarrow 0 \), \( u_0(x) \leftarrow x^{[0]} \), \( v_{-1}(x) \leftarrow x^{[0]} \), \( v_0(x) \leftarrow 0 \);
\While{ \( \deg r_{i-1}(x) \geq d_S \) }
    \State \( q_i(x), r_i(x) \leftarrow RDiv(r_{i-1}(x), r_{i-2}(x)) \);
    \State \( u_i(x) \leftarrow u_{i-2}(x) - q_i(x) \otimes u_{i-1}(x) \);
    \State \( v_i(x) \leftarrow v_{i-2}(x) - q_i(x) \otimes v_{i-1}(x) \);
    \State \( i \leftarrow i + 1 \);
\EndWhile
\State \(\textbf{Output}\): Polynomials $r_{i-1}(x), u_{i-1}(x), v_{i-1}(x)$.
\end{algorithmic}
\end{algorithm}

\begin{definition}
    ($q$-Transform \cite{silva2009fast})
    The $q$-transform of a vector $f\in \mathbb{F}_{q^m}^m$ (or a $q$-polynomial $f(x)$) with respect to a normal element $\alpha$ is the vector $(F_{0},F_{1},\ldots,F_{m-1})$ (or the $q$-polynomial $F(x)$) given by $F_{j}=f(\alpha^{[j]}) = \sum_{i=0}^{m-1}f_{i}\alpha^{[i+j]}$, $j=0,1,\ldots,m-1$.
\end{definition}

\begin{definition}
    (Inverse $q$-Transform \cite{silva2009fast}) The inverse $q$-transform of a vector $F = (F_{0}~ F_{1}~\ldots F_{m-1}) \in \mathbb{F}_{q^m}^{m}$ (or a $q$-polynomial $F(x)$) with respect to $\alpha$ is given by $f_{i} = F(\bar{\alpha}^{[j]})$, $j= 0, \ldots,m-1$, where the bases $\{\bar{\alpha}_{0},\ldots,\bar{\alpha}_{m-1}\}$ is dual to $\{\alpha_{0},\ldots,\alpha_{m-1}\}$.
\end{definition}

 Let $\bm{c}$ be the transmitted codeword corrupted by the error vector $\bm{e}$ of the rank weight $r$. Let $\bm{y}$ be the received vector, and $\bm{y} = \bm{c} + \bm{e}$. Then there exists the error location matrix $Y \in \mathbb{F}_{q}^{r \times n}$ of rank $r$ and the error value vector $\bm{\epsilon} = (\epsilon_{1},\epsilon_{2},\ldots,\epsilon_{r}) \in \mathbb{F}_{q^m}^{r}$ that satisfy $\bm{e = \epsilon} \cdot Y = (\epsilon_{1},\epsilon_{2},\ldots,\epsilon_{r}) \cdot Y$.

Let $y(x)$ be the linearized polynomial corresponding to the received vector 
$\bm{y}$, and let $Y(x)$ be its $q$-transform. Similarly, let $F(x)$ be the $q$-transform of the codeword $f(\bm{g})$. Since $y = f(\bm{g}) +\bm{e}$, we have $Y(x) = F(x)+E(x)$, where $E(x)$ is the $q$-transform of $\bm{e}$. The error locator polynomial $\wedge(x)$ then satisfies the following key equation in the transformed domain.

\begin{theorem}
    (Transformed Key Equation \cite{silva2009fast}) Let $E(x)$ be the $q$-transform of the error word $e(x)$, then the error locator polynomial $\wedge(x)$ satisfies
    \begin{equation}\label{equa2}
        \wedge(x) \otimes E(x) \equiv 0 \mod (x^{[m]}-x),
    \end{equation}    
    where $deg_{q}\wedge(x) = r \leq  \left \lfloor \frac{t-k}{2} \right \rfloor$, $k \leq t \leq min\{m,n\}$.
\end{theorem}

The following Algorithm \ref{alg:alg2} provides a method for calculating the symbolic product $h(x) = f(x) \otimes g(x)= f\Big(g(x)\Big) \mod (x^{[m]}-x)$, where $H(x)$ is $q$-transform of $h(x)$ and $H_{i} = h(\alpha^{[i]})= f\Big(g(\alpha^{[i]})\Big)$.

\begin{algorithm}[h]
\caption{Fast Symbolic Product}
\label{alg:alg2}
\begin{algorithmic}[1]
\State \textbf{Input}: Linearized polynomials \( f(x), g(x) \);
\State Calculate the \(q\)-transform of \(g(x)\): \( G_{i} = g(\alpha^{[i]}) \), for \(i=0,\ldots,m-1\);
\State Calculate \(H_{i} = f(G_{i})\) for \(i=0,\ldots,m-1, H(x) = \sum_{i=0}^{m-1}H_{i}x^{[i]}\);
\State Calculate inverse \(q\)-transform of \(H(x): h_{i} = H(\alpha^{[i]})\) for \(i=0,\ldots,m-1\) ;
\State \textbf{Output}: Symbolic poduct \( h(x)=\sum_{i=0}^{m-1}h_{i}x^{[i]} \).
\end{algorithmic}
\end{algorithm}

We denote $G_{i} = \sum_{j=0}^{m-1}\gamma_{j}^{(i)}\cdot \alpha^{[j]}$, where $\gamma_{j}^{(i)} \in \mathbb{F}_{q}$ for $0 \leq i,j \leq m-1$. All $m$ evaluation values can be calculated by 
\begin{equation}
    \bigg( f(G_{0})~f(G_{1})~\ldots~f(G_{m-1}) \bigg)= \bigg(f(\alpha^{[0]})~f(\alpha^{[1]})~\ldots~f(\alpha^{[m-1]})\bigg) \cdot
    \begin{pmatrix}
    \gamma_{0}^{(0)} & \gamma_{0}^{(1)} & \ldots & \gamma_{0}^{(m-1)} \\
    \gamma_{1}^{(0)} & \gamma_{1}^{(1)} & \ldots & \gamma_{1}^{(m-1)} \\
     \vdots  &  \vdots   & \vdots  &\vdots\\
    \gamma_{m-1}^{(0)} & \gamma_{m-1}^{(1)} & \ldots & \gamma_{m-1}^{(m-1)} \\
    \end{pmatrix}
    \nonumber
\end{equation}

\begin{equation}
\label{equa1}
 =\bigg(\alpha^{[0]}~ \alpha^{[1]}~ \ldots ~\alpha^{[m-1]}\bigg)
 \begin{pmatrix}
   f_{0}^{(0)} & f_{0}^{(1)} & \ldots & f_{0}^{(m-1)} \\
    f_{1}^{(0)} & f_{1}^{(1)} & \ldots & f_{1}^{(m-1)} \\
     \vdots  &  \vdots   & \vdots  &\vdots\\
    f_{m-1}^{(0)} & f_{m-1}^{(1)} & \ldots & f_{m-1}^{(m-1)} \\   
 \end{pmatrix}
    \begin{pmatrix}
    \gamma_{0}^{(0)} & \gamma_{0}^{(1)} & \ldots & \gamma_{0}^{(m-1)} \\
    \gamma_{1}^{(0)} & \gamma_{1}^{(1)} & \ldots & \gamma_{1}^{(m-1)} \\
     \vdots  &  \vdots   & \vdots  &\vdots\\
    \gamma_{m-1}^{(0)} & \gamma_{m-1}^{(1)} & \ldots & \gamma_{m-1}^{(m-1)} \\   
    \end{pmatrix}   
    \overset{\underset{\mathrm{def}}{}}{=} \bm{\alpha \cdot F \cdot G}.
\end{equation}
The Algorithm \ref{alg:alg3} provides a method for calculating multipoint evaluation of a linearized polynomial. 

\begin{algorithm}[h]
    \caption{Multipoint Evaluation of a Linearized Polynomial}
\label{alg:alg3}
\begin{algorithmic}[1]
\State \textbf{Input}: Linearized polynomials \( f(x)\), evaluation points \( G_{0}, \ldots, G_{m-1}\);
\State Calculate the \(q\)-transform of \(f(x)\): \( F_{i} = f(\alpha^{[i]}) \), for \(i=0,\ldots,m-1\);
\State Calculate representation of $F(x)$ over \(\mathbb{F}_{q}: f_{j}^{(i)}, 0 \leq i,j \leq m-1\);
\State Calculate  \(\bf F \cdot G\) ;
\State \textbf{Output}: \( f(G_{0}),\ldots, f(G_{m-1})\).
\end{algorithmic}
\end{algorithm}

It is noted that there exists a matrix $M \in \mathbb{F}_{q}^{m \times m}$ such that Equation (\ref{equa1}) can be rewritten as $$\bigg(f(G_{0})~f(G_{1})~\ldots f(G_{m-1})\bigg) = (\alpha^{[0]},\ldots,\alpha^{[m-1]})M=(\alpha^{[0]},\ldots,\alpha^{[m-1]})\cdot F \cdot G,$$ therefore we have $M=F \cdot G$. For right symbolic division, this means that matrices $M$ and $G$ are known, and we need to solve $M=F \cdot G$ for $F$. For left symbolic division, this means that $M$ and $F$ are known, and we need to solve $M=F \cdot G$ for $G$. Moreover, there exist some polynomial $\theta(x)$ such that Equation (\ref{equa2}) can be rewritten as $\wedge(x) \otimes E(x) =\theta(x) \cdot (x^{[m]}-x)$, where $Y(x) = f(x) + E(x)$, and therefore we have $\wedge(x) \otimes f(x) =\wedge(x) \otimes Y(x)-\theta(x) \cdot (x^{[m]}-x)$. The following Algorithm \ref{alg:alg4} provides a method for solving the transformed key equation with the LEEA.

\begin{algorithm}[h]
    \caption{Solving the Transformed Key Equation with the LEEA}
\label{alg:alg4}
\begin{algorithmic}[1]
\State \textbf{Input}: Received word \( y(x)\);
\State \(y(x),u(x),v(x) \leftarrow LEEA(x^{[m]}-x,Y(x))\) with  \(d_{S}=\left \lfloor \frac{t+k}{2} \right \rfloor\);
\State \(\hat{f(x)},r(x) \leftarrow LDiv(y(x),u(x))\);
\State \textbf{if} \(r(x) = 0\) \textbf{then} 
\State \hspace{1cm}\textbf{Output}: Evaluation Polynomial \( \hat{f(x)},\wedge(x)=u(x)\);
\State \textbf{else}
\State \hspace{1cm}\textbf{Output}: Decoding Failure.
\end{algorithmic}
\end{algorithm}

\begin{remark}
The proposed decoding algorithm directly recovers the codeword polynomial $f(x)$ without first explicitly solving for the error vector $\bm{e}$, simplifying the process. Its computational complexity is dominated by the $q$-transforms and linear algebra operations, requiring $O(m^{3}log~m)$ operations in $\mathbb{F}_{q}$ per decoding instance. The algorithm guarantees successful decoding if and only if the rank weight of the error satisfies $wt_{R}(\bm{e}) \leq \left \lfloor \frac{t-k}{2} \right \rfloor$, which is the classical error-correction capability of the EG code. In contrast to some prior schemes \cite{song2025interleaved}\cite{ref7} that may have a non-zero decoding failure probability even within this bound, our algorithm is deterministic within its capacity. Furthermore, in the cryptosystems we will construct in Section \ref{sec666}, parameters are chosen such that the induced error weight is always below this capability, thereby ensuring zero decryption failure probability for the overall encryption scheme. 
\end{remark}

\subsection{Decoding Extended Gabidulin-Kronecker product codes}
\label{sec333}
In this section, we will give the decoding method of the Extended Gabidulin-Kronecker product codes. Before presenting the decoding algorithm of the Extended Gabidulin-Kronecker product code, we first provide the definition of the information set.

\begin{definition}
    (Information set \cite{sun2024})
    Let $G$ be a generator matrix of code $\mathcal{C}$, and let $I$ be a subset of $\{1,2,\dots,n\}$ with size $k$. The columns of $G$ indexed by $I$ form a $k \times k$ submatrix, which is denoted by $G_{\{\cdot,I\}}$. If $G_{\{\cdot,I\}}$ is invertible, then the set $I$ is called an information set.
\end{definition}

\begin{theorem}
 Given $\bm{g}_{1}=(g_{1},\ldots,g_{n_{1}})\in \mathbb{F}_{q^m}^{n_{1}}$ with $wt_{R}(\bm{g}_{1}) = t_{1}$, and $\bm{g}_{2}=(g_{1}^{'},\ldots,g_{n_{2}} ^{'})\in \mathbb{F}_{q^m}^{n_{2}}$ with $wt_{R}(\bm{g}_{2}) = t_{2}$, and $\bm{y}=(\bm{y}_{1},\ldots,\bm{y}_{n_{1}}) \in \mathbb{F}_{q^m}^{n_{1}n_{2}}$.
 The problem of decoding an $[n_{1}n_{2},k_{1}k_{2}]$  EGK code $DecEGKCode\Big((\bm{g_{1}} \otimes \bm{g_{2}}),\bm{y}\Big)$ can be efficiently reduced to solving $k_{1}$ instances of decoding an $[n_{2},k_{2}]$ EG code, plus solving one linear system over $\mathbb{F}_{q^m}$.  
\end{theorem}

\begin{proof}
 Let $\bm{x} = (\bm{x}_{1},\dots,\bm{x}_{k_{1}}) \in \mathbb{F}_{q^m}^{k_{1}k_{2}}$ be the plaintext, where $\bm{x}_{i} \in \mathbb{F}_{q^m}^{k_{2}}$, for $1 \leq i \leq k_{1}$. Let $\bm{y} \in \mathbb{F}_{q^m}^{n_{1}n_{2}}$ be the receiving vector, $\bm{e} = (\bm{e}_{1},\dots,\bm{e}_{n_{1}}) \in \mathbb{F}_{q^m}^{n_{1}n_{2}}$ be the error vector, $wt_{R}(\bm{e}) \leq r$, and $wt_{R}(\bm{e}_{i}) \leq r$ for $1 \leq i \leq n_{1}$, and satisfy $\bm{y} = \bm{x}G + \bm{e}$, where $G = G_{1} \otimes G_{2}$, $G_1$ and $G_2$ satisfy the aforementioned Definition \ref{EGK}. We rewrite the above equation as follows.

\begin{align}
\nonumber 
    \bm{y} & = (\bm{y}_{1},\ldots,\bm{y}_{n_{1}})  \\
     &  = (\bm{x}_{1},\ldots,\bm{x}_{k_{1}})
     \left[
\begin{array}{ccc}
    g_{11}G_{2} & \ldots & g_{1n_{1}}G_{2} \\
    \vdots & \ddots & \vdots\\
    g_{k_{1}1}G_{2} & \ldots & g_{k_{1}n_{1}}G_{2}
    \nonumber 
\end{array}
 \right] + (\bm{e}_{1},\ldots,\bm{e}_{n_{1}})\\
 & = \bigg(\sum_{i=1}^{k_{1}}\bm{x}_{i}g_{i1}G_{2} + \bm{e}_{1},\ldots,\sum_{i=1}^{k_{1}}\bm{x}_{i}g_{in_{1}}G_{2} + \bm{e}_{n_{1}}\bigg).
 \nonumber 
\end{align}

Therefore we obtain the following system:
\begin{align}
\begin{split}
\left \{
\begin{array}{ll}
    \bm{y}_{1} & = \bigg(\sum_{i=1}^{k_{1}}\bm{x}_{i}g_{i1}\bigg)G_{2} + \bm{e}_{1},\\
    & \vdots\\
   \bm{y}_{n_{1}} & = \bigg(\sum_{i=1}^{k_{1}}\bm{x}_{i}g_{in_{1}}\bigg)G_{2} + \bm{e}_{n_{1}}.
\end{array}
\right. 
\end{split}
\nonumber
\end{align}

Arbitrarily selecting an information set $I=\{j_{1},\ldots,j_{k_{1}}\} \in \{1,2,\ldots,n_{1}\}$ of $G_1$, we obtain: 
\begin{align} \label{222}
\begin{split}
\left \{
\begin{array}{ll}
    \bm{y}_{j_1} & = \bigg(\sum_{i=1}^{k_{1}}\bm{x}_{i}g_{ij_{1}}\bigg)G_{2} + \bm{e}_{j_1} \overset{\underset{\mathrm{def}}{}}{=} \bm{c}_{1} + \bm{e}_{j_{1}},\\
    & \vdots\\
   \bm{y}_{j_{k_{1}}} & = \bigg(\sum_{i=1}^{k_{1}}\bm{x}_{i}g_{ij_{k_{1}}}\bigg)G_{2} + \bm{e}_{j_{k_{1}}} \overset{\underset{\mathrm{def}}{}}{=} \bm{c}_{k_{1}} + \bm{e}_{j_{k_1}}.
\end{array}
\right. 
\end{split}
\end{align}   

Furthermore, the system of equations can be rewritten as: \\
\begin{equation}
\nonumber
\begin{bmatrix}
\bm{y}_{j_1}^{T} \\
\bm{y}_{j_2}^{T} \\
\vdots \\
\bm{y}_{j_{k_1}}^{T}
\end{bmatrix}
=
\begin{bmatrix}
g_{1j_1} G_{2}^{T} & g_{1j_2} G_{2}^{T} & \cdots & g_{1j_{k_1}} G_{2}^{T} \\
g_{2j_1} G_{2}^{T} & g_{2j_2} G_{2}^{T} & \cdots & g_{2j_{k_1}} G_{2}^{T}\\
\vdots & \vdots & \ddots & \vdots \\
g_{k_1 j_1} G_{2}^{T} & g_{k_1 j_2} G_{2}^{T} & \cdots & g_{k_1 j_{k_1}} G_{2}^{T}
\end{bmatrix}
\cdot
\begin{bmatrix}
\bm{x}_{1}^{T} \\
\bm{x}_{2}^{T} \\
\vdots \\
\bm{x}_{k_1}^{T}
\end{bmatrix}
+ 
\begin{bmatrix}
\bm{e}_{j_{1}}^{T} \\
\bm{e}_{j_{2}}^{T}\\
\vdots \\
\bm{e}_{j_{k_{1}}}^{T}
\end{bmatrix}
\end{equation}

To recover the plaintext information, we need to solve the aforementioned Equation (\ref{222}). Each equation in this system is of the form $\bm{y}_{j_{s}} = \bm{c}_{s} + \bm{e}_{j_{s}}$, where $\bm{c}_{s}$ is an unknown codeword of the EG code $\mathcal{C}_{2}$. Thus, decoding each $\bm{y}_{j_{s}}$ (for $s=1,\ldots,k_{1}$) as an instance of $DecEGCode(\bm{g}_{2},\bm{y}_{j_{s}})$ yields candidates for $\bm{c}_{s}$. Once all $\bm{c}_{s}$ are recovered, the original message $\bm{x}$ can be retrieved by solving a linear system derived from the relation between the $\bm{c}_{s}$  and $\bm{x}$.    $\hfill\blacksquare$
\end{proof}

It can be seen from the above proof process that we are not decoding codes with length $n$, but rather dividing the code into $n_{1}$ blocks with smaller length $n_{2}$, which results in smaller decoding complexity. Specifically, we first need to select $k_{1}$ equations with $k_{1}$ unknowns from an information set of $G_{1}$, with a cost of $\mathcal{O}(n_{2}^{3}k_{1}^{3}k_{2}m^{2})$ over $\mathbb{F}_{q}$. And then decode each equation, which means decoding EG codes, as has been detailed in Section \ref{sec3}. After obtaining the solutions $\bm{c}_{1},\ldots,\bm{c}_{k_{1}}$ of the decoded EG code, and given that $G_1$ and $G_2$ are known, the plaintext message can be obtained by solving the system of linear equations with $k_{1}k_{2}$ unknowns and $k_{1}n_{2}$ linear equations.

Next, we will present the bounds on the error-correcting capability of Extended Gabidulin-Kronecker product codes. Prior to this, Reference \cite{song2025interleaved} has given the following bounds on the error-correcting capability of EG codes.

\begin{theorem} \cite{song2025interleaved}
    The $EG_{k}(g)$ codes with parameters $[n,k,d]$ and with a generator of weight $t$ can decode errors of weight up to $min\Big\{t-k,\left \lfloor \frac{n-k}{2}\right \rfloor\Big\}$.
\end{theorem}

\begin{theorem} 
  The Extended Gabidulin-Kronecker product codes defined in Definition \ref{EGK} can correct errors of weight $r \leq \left \lfloor \frac{t_{2}-k_{2}}{2} \right \rfloor$ for the case of $k_{1}=t_{1},t_{2}=m<t_{1}t_{2}$. And for the case of $k_{1}\leq t_{1},k_{2}\leq t_{2},t_{1}t_{2}\leq m$, the code $\mathcal{C}$ can correct errors of weight $r \leq \frac{(t_{1}-k_{1}+1)(t_{2}-k_{2}+1)-1}{2}$. 
\end{theorem}
\begin{proof}
  Let $d$ be the minimum rank distance of the EGK code $\mathcal{C}$. According to the sphere packing principle, if the error-correcting spheres (rank weight spheres centered at the codewords with a specific radius) of two distinct codewords do not intersect, unique error correction is achievable. For any two distinct codewords $\bm{c}_{1},\bm{c}_{2}\in \mathcal{C}$, the rank-distance spheres of radius $r$ centered at them are disjoint if $d>2r$. This is because if there existed a vector $\bm{y}$ such that $d_{R}(\bm{c}_{1},\bm{y}) \leq r$, and $d_{R}(\bm{c}_{2},\bm{y})\leq r$, then by the triangle inequality, $d_{R}(\bm{c}_{1},\bm{c}_{2})\leq d_{R}(\bm{c}_{1},\bm{y})+d_{R}(\bm{y},\bm{c}_{2})$, contradicting $d_{R}(\bm{c}_{1},c_{2}) \ge d$. Therefore, unique decoding is guaranteed for any error of weight $r \leq \left \lfloor \frac{d-1}{2} \right \rfloor$. Substituting the values of $d$ from Proposition \ref{minimal distance} for the two parameter regimes yields $r \leq \left \lfloor \frac{t_{2}-k_{2}}{2} \right \rfloor$ for the case of $k_{1}=t_{1},t_{2}=m<t_{1}t_{2}$. And for the case of $k_{1}\leq t_{1},k_{2}\leq t_{2},t_{1}t_{2}\leq m$, the code $\mathcal{C}$ can correct errors of weight $r \leq \left \lfloor \frac{d-1}{2} \right \rfloor \leq \frac{d-1}{2} \leq \frac{(t_{1}-k_{1}+1)(t_{2}-k_{2}+1)-1}{2}$.

From the above, it can be concluded that code $\mathcal{C}$ can correct errors of weight $r \leq \left \lfloor \frac{t_{2}-k_{2}}{2} \right \rfloor$ for the case of $k_{1}=t_{1},t_{2}=m<t_{1}t_{2}$. And for the case of $k_{1}\leq t_{1},k_{2}\leq t_{2},t_{1}t_{2}\leq m$, the code $\mathcal{C}$ can correct errors of weight $r \leq \frac{(t_{1}-k_{1}+1)(t_{2}-k_{2}+1)-1}{2}$.
\end{proof} $\hfill\blacksquare$

\begin{remark}
 The overall decoding complexity comprises two parts: (i) solving $k_{1}$ instances of EG code decoding, requiring $\mathcal{O}(k_{1}m^{3}log~m)$ operations over $\mathbb{F}_{q}$; and (ii) solving a linear system of size $\mathcal{O}(k_{1}k_{2}) \times \mathcal{O}(k_{1}n_{2})$ over $\mathbb{F}_{q^m}$, which via Gaussian elimination costs
$\mathcal{O}((k_{1}k_{2})^{2}\cdot k_{1}n_{2} \cdot m)=\mathcal{O}(k_{1}^{3}k_{2}^{2}n_{2}m^{2})$ operations over $\mathbb{F}_{q}$ (since each $\mathbb{F}_{q^m}$ operation costs $\mathcal{O}(m)$  $\mathbb{F}_{q}$-operations). In typical cryptographic parameter settings where $m$ is the largest scaling parameter, the first term $\mathcal{O}(k_{1}m^{3}log~m)$ dominates the complexity. Furthermore, the blockwise structure of the decoding algorithm may, under certain favorable distributions of the error, allow for the correction of error patterns whose total weight exceeds the classical bound 
 $\left \lfloor \frac{d-1}{2} \right \rfloor$, although this is not guaranteed. This is because during the decoding process, each block can correct $r$ errors, which increases the error-correcting capability of code $\mathcal{C}$ up to $n_{1}r$. In general, this value will exceed the error-correcting capability of code $\mathcal{C}$. In our cryptographic constructions, parameters are chosen such that the error weight is strictly within the guaranteed correction capability, ensuring deterministic decryption.
\end{remark}

\section{Applications to cryptography}
\label{sec666}
\subsection{RQC from the EGK codes with blockwise errors}
\subsubsection{Protocol Description}
\label{secA}
Our first new encryption scheme stands for the improved RQC with unique syndrome and uses the following three codes:

$\star$ A public $[n,k]_{q^m}$-EGK code with generator matrix $G \in \mathbb{F}_{q^m}^{k_{1}k_{2}\times n_{1}n_{2}}$ generated by $\bm{g}_{1}\in \mathbb{F}_{q^m}^{n_{1}}$ with weight $t_{1}$ and $\bm{g}_{2}\in \mathbb{F}_{q^m}^{n_{2}}$ with weight $t_{2}$. This code can correct errors of weight $r \leq \left \lfloor \frac{t_{2}-k_{2}}{2} \right \rfloor$ for the case of $k_{1}=t_{1},t_{2}=m<t_{1}t_{2}$, and for the case of $k_{1}\leq t_{1},k_{2}\leq t_{2},t_{1}t_{2}\leq m$, $r \leq  \frac{(t_{1}-k_{1}+1)(t_{2}-k_{2}+1)-1}{2}$ in a deterministic way. 

$\star$ A random $[2n,n]_{q^m}$-ideal code, of parity matrix $[I_{n}~ ~~\mathcal{IM}(\bm{h})^{T}]$.

$\star$ A random $[3n,n]_{q^m}$-ideal code, of parity matrix 
$\begin{bmatrix}
I_{n}      & \bm{0} & \mathcal{IM}(\bm{h})^{T}      \\
\bm{0}      & I_{n} & \mathcal{IM}(\bm{s})^{T}
\end{bmatrix}.
$

The four polynomial-time algorithms constituting our scheme are illustrated in Fig.\ref{fig:placeholder}. The generator matrix of the EGK code is publicly known, it is only used in the decryption step and does not contribute to the security of the scheme. We use two random codes to ensure the security of the scheme.

\begin{figure}[h]
\begin{minipage}{\linewidth}
\begin{center}
    \fbox{ 
        \begin{minipage}{0.9\textwidth} 
            \underline{RQC.Setup($1^{\lambda}$):}\\
            \hspace{2em} $\bullet$ Generates and outputs the global parameters $param = (n,k,n_{1},n_{2},k_{1},k_{2},t_{1},t_{2},\omega_{\bm{x}},\omega_{\bm{y}},\omega_{1},\omega_{\bm{e}},\omega_{2},r)$;\\
            \underline{RQC.KGen($\lambda$):}\\
            \hspace{1cm} $\bullet$ Sample ${\bm{x},\bm{y}}\stackrel{\text{\$}}{\leftarrow}\mathcal{S}_{\omega_{\bm{x}},\omega_{\bm{y}}}^{n,n}$ from a seed $seed_{1}$ of 40 bytes;\\
            \hspace{2em} $\bullet$ Sample $\bm{g}_{1} \stackrel{\text{\$}}{\leftarrow} \mathcal{S}_{t_{1}}^{n_{1}}(\mathbb{F}_{q^m})$, $\bm{g}_{2} \stackrel{\text{\$}}{\leftarrow} \mathcal{S}_{t_{2}}^{n_{2}}(\mathbb{F}_{q^m})$, $\bm{h} \stackrel{\text{\$}}{\leftarrow} \mathbb{F}_{q^m}^{n}$ from a seed $seed_{2}$ of 40 bytes; \\
            \hspace{2em} $\bullet$ Compute $\bm{s}=\bm{x}+\bm{hy} ~ mod ~N(X)$;\\
            \hspace{2em} $\bullet$ Output a public key $pk = (\bm{g_{1}},\bm{g_{2}},\bm{h},\bm{s})$ and a private key $sk = (\bm{x},\bm{y})$.\\
            \underline{RQC.Enc(pk, $\bm{m}$):}\\
             \hspace{2em} $\bullet$ Input the public key $pk$ and a plaintext $\bm{m} \in \mathbb{F}_{q^m}^{k}$;\\
        \hspace{2em} $\bullet$ Compute the generator matrix $G_{1} \in \mathbb{F}_{q^m}^{k_{1} \times n_{1}}$ of the $[n_{1},k_{1}]_{q^m}$ Extended Gabidulin codes defined by $\bm{g}_{1}$;\\
              \hspace{2em} $\bullet$  Compute the generator matrix $G_{2} \in \mathbb{F}_{q^m}^{k_{2} \times n_{2}}$ of the $[n_{2},k_{2}]_{q^m}$ Extended Gabidulin codes defined by $\bm{g}_{2}$;\\
               \hspace{2em} $\bullet$ Compute the generator matrix $G = G_{1} \otimes G_{2} \in \mathbb{F}_{q^m}^{k \times n}$ of the $[n =n_{1}n_{2}, k= k_{1}k_{2}]_{q^m}$ Extended Gabidulin-Kronecker product codes defined by $\bm{g}_{1},\bm{g}_{2}$;\\
               \hspace{2em} $\bullet$ Use the randomness $\theta$ to generate $(\bm{r}_{1},\bm{e},\bm{r}_{2}) \stackrel{\text{\$}}{\leftarrow} \mathcal{S}_{\omega_{1},\omega_{\bm{e}},\omega_{2}}^{n,n,n}$;\\
              \hspace{2em} $\bullet$ Compute $\bm{u}=\bm{r}_{1}+\bm{h} \cdotp \bm{r}_{2} ~ mod ~N(X)$ and $\bm{v} = \bm{m} \cdotp G+\bm{s} \cdotp \bm{r}_{2}+\bm{e}~ mod ~N(X)$;  \\
              \hspace{2em} $\bullet$  Output a ciphertext $\bm{c} = (\bm{u},\bm{v})$.  \\
            \underline{RQC.Dec(sk, $\bm{c}$):}\\
             \hspace{2em} $\bullet$ Input the private key $sk$ and the ciphertext $\bm{c}$;\\
             \hspace{2em} $\bullet$ Output the plaintext $\bm{m} := DecEKGCode\Big((\bm{g}_{1}\otimes\bm{g}_{2}),\bm{v}-\bm{y} \cdotp \bm{u}\Big)$.
        \end{minipage} }
\end{center}
\end{minipage}
 \caption{Description of our proposal RQC.EGK-BWE}
    \label{fig:placeholder}
\end{figure}

\textbf{Correctness.}  The correctness of our encryption scheme clearly relies on the decoding capability of the Extended Gabidulin-Kronecker product code $\mathcal{C}$. Specifically, assume that code $\mathcal{C}$ can correctly decode $\bm{v}-\bm{y \cdot u}$, then $$Dec(sk,Enc(\bm{m},pk)) = \bm{m}$$
holds. Meanwhile, for code $\mathcal{C}$ to correctly decode $\bm{v}-\bm{y \cdot u}$, we have 
\begin{align*}
\nonumber
    \bm{v-y} \cdot \bm{u} &= \bm{m}G + \bm{s}\cdot \bm{r}_{2}+\bm{e}-\bm{y}\cdot \bm{u}\\  
    &= \bm{m}G + (\bm{x}+\bm{h}\cdot \bm{y})\cdot \bm{r}_{2}+\bm{e}-\bm{y}(\bm{r}_{1}+\bm{h} \cdot \bm{r}_{2}) \\ 
    &= \bm{m}G + (\bm{x} \cdot \bm{r}_{2}-\bm{y}\cdot \bm{r}_{1}+\bm{e}).
\end{align*}
The error term $\bm{x} \cdot \bm{r}_{2}-\bm{y}\cdot \bm{r}_{1}+\bm{e}$ must satisfy
\begin{equation*}
  wt_{R}(\bm{x}\cdot \bm{r}_{2}-\bm{y} \cdot \bm{r}_{1}+\bm{e}) \leq \omega_{\bm{x}}\omega_{2}+w_{\bm{y}}\omega_{1}+\omega_{\bm{e}}.   
\end{equation*}
Let $r = \omega_{\bm{x}}\omega_{2}+w_{\bm{y}}\omega_{1}+\omega_{\bm{e}}$. For correct decryption, we must ensure $r \leq \left \lfloor \frac{t_{2}-k_{2}}{2} \right \rfloor$ for the case of $k_{1}=t_{1},t_{2}=m<t_{1}t_{2}$, and for the case of $k_{1}\leq t_{1},k_{2}\leq t_{2},t_{1}t_{2}\leq m$, $r \leq  \frac{(t_{1}-k_{1}+1)(t_{2}-k_{2}+1)-1}{2}$, and the message $\bm{m}$ can be correctly decoded. 

The decoding algorithm described in Section \ref{sec333} is deterministic, meaning it has no decryption failure probability, or equivalently, the occurrence of decryption failure is null.

\subsubsection{KEM version}
To obtain a more secure encryption scheme, this section transforms the aforementioned proposed scheme into an IND-CCA2 secure KEM scheme, we adopt the technique proposed in \cite{ref5}. Our KEM scheme is tight, and the KEM version of the new scheme is illustrated in Fig.\ref{fig:555}. Among them, $\mathcal{G}$, $\mathcal{H}$, and $\mathcal{K}$ are pairwise distinct hash functions, typically SHA512 recommended by NIST, $\eth$ denotes the instance of the new improved RQC cryptosystem as described above.
\begin{figure}[h]
\begin{minipage}{\linewidth}
\begin{center}
    \fbox{ 
        \begin{minipage}{0.9\textwidth} 
            \underline{Setup($1^{\lambda}$):}\\
            \hspace{2em} $\bullet$ Generates and outputs the global parameters $param = (n,k,n_{1},n_{2},k_{1},k_{2},t_{1},t_{2},\omega_{\bm{x}},\omega_{\bm{y}},\omega_{1},\omega_{\bm{e}},\omega_{2},r)$;\\
            \hspace{2em} $\bullet$ The plaintext space has size $k_{1} \times k_{2} \times m > 256$ as required by NIST.\\
            \underline{KGen(param):}\\
            \hspace{1cm} $\bullet$ Sample ${\bm{x},\bm{y}}\stackrel{\text{\$}}{\leftarrow}\mathcal{S}_{\omega_{\bm{x}},\omega_{\bm{y}}}^{n,n}$ from a seed $seed_{1}$ of 40 bytes;\\
            \hspace{2em} $\bullet$ Sample $\bm{g}_{1} \stackrel{\text{\$}}{\leftarrow} \mathcal{S}_{t_{1}}^{n_{1}}(\mathbb{F}_{q^m})$, $\bm{g}_{2} \stackrel{\text{\$}}{\leftarrow} \mathcal{S}_{t_{2}}^{n_{2}}(\mathbb{F}_{q^m})$, $\bm{h} \stackrel{\text{\$}}{\leftarrow} \mathbb{F}_{q^m}^{n}$ from a seed $seed_{2}$ of 40 bytes; \\
            \hspace{2em} $\bullet$ Compute $\bm{s}=\bm{x}+\bm{hy} ~ mod ~N(X)$;\\
            \hspace{2em} $\bullet$ Output a public key $pk = (\bm{g_{1}},\bm{g_{2}},\bm{h},\bm{s})$ and a private key $sk = (\bm{x},\bm{y})$.\\
            \underline{Encapsulate(pk):}\\
             \hspace{2em} $\bullet$ Randomly generate $\bm{m} \stackrel{\text{\$}}{\leftarrow} \mathbb{F}_{q^m}^{k_{1}k_{2}}$, and obtain a random element $\theta \leftarrow \mathcal{G}(\bm{m})$; \\
             \hspace{2em} $\bullet$ Generate the ciphertext $\bm{c} \leftarrow (\bm{u},\bm{v})= \eth.Encrypt(pk,\bm{m},\theta)$; \\
             \hspace{2em} $\bullet$  Obtain the symmetric key $K \leftarrow \mathcal{K}(\bm{m},\bm{c})$;\\
             \hspace{2em} $\bullet$ Let $d \leftarrow \mathcal{H}(\bm{m})$, and send $(\bm{c},\bm{d})$.\\
            \underline{Decapsulate(sk,$\bm{c}$,$\bm{d}$):}\\
             \hspace{2em} $\bullet$ Decrypt $\bm{m}^{'} \leftarrow \eth.Decrypt(sk,\bm{c})$ and compute $\theta^{'} \leftarrow \mathcal{G}(\bm{m})$.\\
             \hspace{2em} $\bullet$ (Re-)encrypt $\bm{m}^{'}$, and obtain $\bm{c}^{'} \leftarrow \eth.Encrypt(pk,\bm{m}^{'},\theta^{'})$.\\
             \hspace{2em} $\bullet$ If $\bm{c} \neq \bm{c}^{'}$ or $\bm{d} \neq \mathcal{H}(\bm{m})$, then abort. Otherwise, obtain the shared key $K \leftarrow \mathcal{K}(\bm{m},\bm{c})$.
        \end{minipage} }
\end{center}
\end{minipage}
 \caption{Description of our proposal RQC.EGK-BWE KEM}
    \label{fig:555}
\end{figure}
 \begin{theorem}
     The scheme RQC.EGK-BWE presented in Sec.\ref{secA} is IND-CPA secure under the 2-DIBRSD and 3-DIBRSD problems. To be specific, for any probabilistic polynomial-time adversary $\mathcal{A}$, there exists a probabilistic polynomial-time adversary $\mathcal{B}$ such that the advantage of $\mathcal{A}$ against the IND-CPA experiment in our improved RQC is bounded by: $$Adv_{RQC.EGK-BWE, \mathcal{A}}^{IND-CPA}(\lambda) \leq 2\left(Adv_{\mathcal{B}}^{2-DIBRSD(q,m,n,r,\bm{\eta}_{2},\bm{\rho}_{2})}(\lambda)+Adv_{\mathcal{B}}^{3-DIBRSD(q,m,n,r^{'},\bm{\eta}_{3},\bm{\rho}_{3})}(\lambda)\right).$$
 \end{theorem}
\begin{proof}
   The proof of the improved RQC scheme is similar to the proof from \cite{ref6} with  a $2-DIBRSD(q,m,n,r,\bm{\eta}_{2},\bm{\rho}_{2})$ instance defined from a $[2n,n]$ code, where $\bm{\eta}_{2} = (n,n)$, $\bm{\rho}_{2} = (\omega_{\bm{x}},\omega_{\bm{y}})$, $r= \omega_{\bm{x}}+\omega_{\bm{y}}$, and a $3-DIBRSD(q,m,n,r^{'},\bm{\eta}_{3},\bm{\rho}_{3})$ instance defined from an $[3n,n]$ code, where $\bm{\eta}_{3} = (n,n,n)$, $\bm{\rho}_{3} = (\omega_{\bm{x}},\omega_{\bm{e}},\omega_{\bm{y}})$, $r^{'}=\omega_{\bm{x}}+\omega_{\bm{e}}+\omega_{\bm{y}}$. Two instances are defined as 

\begin{align*}
       \begin{bmatrix} 
       I_n & \mathcal{I}M(\bm{h})^\top 
       \end{bmatrix}
       \begin{bmatrix} 
       \bm{x}^\top \\ \bm{y}^\top 
       \end{bmatrix} &= \bm{s}^\top, \\ 
\begin{bmatrix}
I_n & \bm{0} & \mathcal{I}M(\bm{h})^\top \\ \bm{0} & I_n & \mathcal{I}M(\bm{s})^\top 
\end{bmatrix} 
\begin{bmatrix}
\bm{r}_1^\top \\ \bm{r}_2^\top \\ \bm{e}^\top 
\end{bmatrix}
&= \begin{bmatrix} 
\bm{u}^\top \\ \bm{v}^\top - (\bm{m}G)^\top 
\end{bmatrix}.
\end{align*} $\hfill \blacksquare$
\end{proof}

\subsection{RQC from the EGK codes with non-homogeneous errors}
$\textbf{Protocol Description}$:
\label{rr}
Before presenting our new scheme, we first need to introduce the following notations.
\begin{align*}
    \mathcal{S}_{(\omega_{1},\omega_{2})}^{a \times (b,c,d)} =  \{\bm{X}=(\bm{X}_{1},\bm{X}_{2},\bm{X}_{3})\in \mathbb{F}_{q^m}^{a \times (b,c,d)}\Big| wt_{R}(\bm{X}_{1},\bm{X}_{3})=\omega_{1},wt_{R}(\bm{X}_{2})=\omega_{2},Supp(\bm{X}_{1},\bm{X}_{3}) \subset Supp(\bm{X}_{2})\}.
\end{align*}

Let $\bm{z} = (z_{1},\ldots,z_{a}) \in \mathbb{F}_{q^m}^{a}$, and let $\bm{M} \in \mathbb{F}_{q^m}^{a \times b}$ be a matrix with columns $\bm{M}_{1},\ldots, \bm{M}_{b}$. We define their product $\bm{z} \cdot \bm{M}$ as the matrix whose $j$-th column (for $1 \leq j \leq b$) is the vector $z \cdot M_{j}$, where $\cdot$ denotes the product in the ring $\mathbb{F}_{q^m}[X]/\langle N(X)\rangle$ as defined in Section \ref{sec222}. That is,
\begin{equation}
    \bm{z} \cdot \bm{M} \overset{\underset{\mathrm{\triangle}}{}}{=} \bigg(\Big(\bm{z} \cdot \bm{M}_{1}^{T} mod N(X)\Big)^{T}, \dots, \Big(\bm{z} \cdot \bm{M}_{b}^{T} mod N(X)\Big)^{T}\bigg) \in \mathbb{F}_{q^m}^{a \times b}.
    \nonumber
\end{equation}

    For a integer $a$, let $\bm{v} = (\bm{v_{1}},\bm{v}_{2},\ldots,\bm{v}_{b}) \in \mathbb{F}_{q^m}^{ab}$, where each $\bm{v}_{i} \in \mathbb{F}_{q^m}^{a}$, for $1\leq i \leq b$. The $Fold()$ function reshapes this vector into a matrix: $Fold(\bm{v}) = (\bm{v}_{1}^{T},\bm{v}_{2}^{T},\ldots,\bm{v}_{b}^{T}) \in \mathbb{F}_{q^m}^{a \times b}$. Its inverse is $Unfold()$, which concatenates the columns of a matrix back into a long vector: $Unfold(V)=(V_{\cdot,1}^{T},\ldots,V_{\cdot,b}^{T}) \in \mathbb{F}_{q^m}^{ab}$, where $V_{\cdot,j}$ denotes the $j$-th column of V.
 
Our second new encryption scheme with multiple syndromes and uses the following two codes:

$\star$ An $[n,k]_{q^m}$-EGK code $G \in \mathbb{F}_{q^{m}}^{k_{1}k_{2} \times n_{1}n_{2}}$ generated by $\bm{g}_{1} \in \mathbb{F}_{q^m}^{n_{1}}$ with weight $t_{1}$ and $\bm{g}_{2} \in \mathbb{F}_{q^m}^{n_{2}}$ with weight $t_2$. This code can correct errors of weight $r \leq \left \lfloor \frac{t_{2}-k_{2}}{2} \right \rfloor$ for the case of $k_{1}=t_{1},t_{2}=m<t_{1}t_{2}$, and for the case of $k_{1}\leq t_{1},k_{2}\leq t_{2},t_{1}t_{2}\leq m$, $r \leq  \frac{(t_{1}-k_{1}+1)(t_{2}-k_{2}+1)-1}{2}$ in a deterministic way. 

$\star$  A random $[2n_{2},n_{2}]_{\mathbb{F}_{q^m}}$-ideal code, of parity matrix $[I_{n} ~ \mathcal{IM}(\bm{h})^{T}]$.

 Next, our proposed RQC.EGK-Multi-NH is presented in Figure \ref{fig:111}.

\begin{figure}[htbp]
\begin{minipage}{\linewidth}
\begin{center}
    \fbox{ 
        \begin{minipage}{0.9\textwidth} 
            \underline{RQC.Setup($1^{\lambda}$):}\\
            \hspace{2em} $\bullet$ Generates and outputs the global parameters $param = (n_{1},n_{2},k_{1},k_{2},t_{1},t_{2},\omega_{\bm{x}},\omega_{\bm{y}},\omega_{1},\omega_{2},N(X))$.\\
            \underline{RQC.KGen($\lambda$):}\\
            \hspace{1cm} $\bullet$ Sample ${\bm{x},\bm{y}}\stackrel{\text{\$}}{\leftarrow}\mathcal{S}_{\omega_{\bm{x}},\omega_{\bm{y}}}^{n_{2},n_{2}}$;\\
            \hspace{2em} $\bullet$ Sample $\bm{g}_{1} \stackrel{\text{\$}}{\leftarrow} \mathcal{S}_{t_{1}}^{n_{1}}(\mathbb{F}_{q^m})$, $\bm{g}_{2} \stackrel{\text{\$}}{\leftarrow} \mathcal{S}_{t_{2}}^{n_{2}}(\mathbb{F}_{q^m})$, $\bm{h} \stackrel{\text{\$}}{\leftarrow} \mathbb{F}_{q^m}^{n_{2}}$; \\
            \hspace{2em} $\bullet$ Compute $\bm{s}=\bm{x}+\bm{hy} ~ mod ~N(X)$;\\
            \hspace{2em} $\bullet$ Output a public key $pk = (\bm{g_{1}},\bm{g_{2}},\bm{h},\bm{s})$ and a private key $sk = (\bm{x},\bm{y})$.\\
            \underline{RQC.Enc(pk, $\bm{m}$):}\\
             \hspace{2em} $\bullet$ Input the public key $pk$ and a plaintext $\bm{m} \in \mathbb{F}_{q^m}^{k}$;\\
             \hspace{2em} $\bullet$ Compute the generator matrix $G_{1} \in \mathbb{F}_{q^m}^{k_{1} \times n_{1}}$ of the $[n_{1},k_{1}]_{q^m}$ Extended Gabidulin codes defined by $\bm{g}_{1}$;\\
              \hspace{2em} $\bullet$  Compute the generator matrix $G_{2} \in \mathbb{F}_{q^m}^{k_{2} \times n_{2}}$ of the $[n_{2},k_{2}]_{q^m}$ Extended Gabidulin codes defined by $\bm{g}_{2}$;\\
               \hspace{2em} $\bullet$ Compute the generator matrix $G = G_{1} \otimes G_{2} \in \mathbb{F}_{q^m}^{k \times n}$ of the $[n =n_{1}n_{2}, k= k_{1}k_{2}]_{q^m}$ Extended Gabidulin-Kronecker product codes defined by $\bm{g}_{1},\bm{g}_{2}$;\\
               \hspace{2em} $\bullet$ Use the randomness $\theta$ to generate $(\bm{R}_{1},\bm{E},\bm{R}_{2}) \stackrel{\text{\$}}{\leftarrow} \mathcal{S}_{(\omega_{1},\omega_{2})}^{n_{2} \times (n_{1},n_{1},n_{1})}$;\\
              \hspace{2em} $\bullet$ Compute $\bm{U}=\bm{R}_{1}+\bm{h} \cdotp \bm{R}_{2} ~ mod ~N(X)$ and $\bm{V} = Fold(\bm{m} \cdotp G)+\bm{s} \cdotp \bm{R}_{2}+\bm{E}~ mod ~N(X)$;  \\
              \hspace{2em} $\bullet$  Output a ciphertext $\bm{c} = (\bm{U},\bm{V})$.  \\
            \underline{RQC.Dec(sk, $\bm{c}$):}\\
             \hspace{2em} $\bullet$ Input the private key $sk$ and the ciphertext $\bm{c}$;\\
             \hspace{2em} $\bullet$ Output the plaintext $\bm{m} := DecEKGCode\bigg((\bm{g}_{1}\otimes\bm{g}_{2}),\bm{V}-\bm{y} \cdotp \bm{U}\bigg)$.
        \end{minipage} }
\end{center}
\end{minipage}
 \caption{Description of our proposal RQC.EGK-Multi-NH}
    \label{fig:111}
\end{figure}

$\textbf{Correctness}$:  The correctness of our encryption scheme clearly relies on the decoding capability of the Extended Gabidulin-Kronecker product code $\mathcal{C}$. Specifically, assume that code $\mathcal{C}$ can correctly decode $\bm{V}-\bm{y \cdot U}$, then $$Dec\Big(sk,Enc(\bm{m},pk)\Big) = \bm{m}$$
holds. Meanwhile, for code $\mathcal{C}$ to correctly decode $\bm{V}-\bm{y \cdot U}$, we have 
\begin{align*}
\nonumber
    \bm{V-y} \cdot \bm{U} &= Fold(\bm{m} \cdotp G) + \bm{s}\bm{R}_{2}+\bm{E}-\bm{y}\cdot \bm{U}\\  
    &= Fold(\bm{m}\cdotp G) + (\bm{x}+\bm{h}\cdot \bm{y})\cdot \bm{R}_{2}+\bm{E}-\bm{y}(\bm{R}_{1}+\bm{h} \cdot \bm{R}_{2}) \\ 
    &= Fold(\bm{m}\cdotp G) + (\bm{x} \cdot \bm{R}_{2}-\bm{y}\cdot \bm{R}_{1}+\bm{E}).
\end{align*}
Therefore, we have $Unfold(\bm{V}-y \cdot \bm{U})= \bm{m}G+ Unfold(\bm{x} \cdot \bm{R}_{2}-\bm{y} \cdot \bm{R}_{1}+ \bm{E}) \in \mathbb{F}_{q^m}^{n}$. The error term $Unfold(\bm{x} \cdot \bm{R}_{2}-\bm{y} \cdot \bm{R}_{1}+\bm{E})$ must satisfy
\begin{equation*}
  wt_{R}(Unfold(\bm{x} \cdot \bm{R}_{2}-\bm{y} \cdot \bm{R}_{1}+\bm{E})) \leq r.   
\end{equation*}
where $r \leq \left \lfloor \frac{t_{2}-k_{2}}{2} \right \rfloor$ for the case of $k_{1}=t_{1},t_{2}=m<t_{1}t_{2}$, and for the case of $k_{1}\leq t_{1},k_{2}\leq t_{2},t_{1}t_{2}\leq m$, $r \leq  \frac{(t_{1}-k_{1}+1)(t_{2}-k_{2}+1)-1}{2}$, and the message $\bm{m}$ can be correctly decoded. 

\begin{theorem}
     The scheme RQC.EGK-Multi-NH presented in Sec.\ref{rr} is IND-CPA secure under the DIRSD and DNHRSL problems. To be specific, for any probabilistic polynomial-time adversary $\mathcal{A}$, there exists a probabilistic polynomial-time adversary $\mathcal{B}$ such that the advantage of $\mathcal{A}$ against the IND-CPA experiment in our improved RQC is bounded by:
     $$Adv_{RQC.EGK-Multi-NH, \mathcal{A}}^{IND-CPA}(\lambda) \leq 2\Big(Adv_{\mathcal{B}}^{2-DIRSD(q,m,2n_{2},n_{2},\omega_{\bm{x}}+\omega_{\bm{y}})}(\lambda)+Adv_{\mathcal{B}}^{DNHRSL(q,m,\bm{\eta}_{3},n_{2},\bm{\rho}_{2},n_{1})}(\lambda)\Big).$$
\end{theorem}
\begin{proof}
The proof of the RQC.EGK-Multi-NH scheme is similar to the proof from \cite{ref6} with  a $2-IRSD(q,m,2n_{2},n_{2},\omega_{\bm{x}}+\omega_{\bm{y}})$ instance defined from a $[2n_{2},n_{2}]$ code, and a $NHIRSL(q,m,\bm{\eta}_{3},n_{2},\bm{\rho}_{2},n_{1})$ instance defined from an $[3n_{2},n_{2}]$ code, where $\bm{\eta}_{3}=(n_{2},n_{2},n_{2})$, $\bm{\rho}_{2}=(\omega_{1},\omega_{2})$. Two instances are defined as 

\begin{align*}
       \begin{bmatrix} 
       I_{n_{2}} & \mathcal{I}M(\bm{h})^\top 
       \end{bmatrix}
       \begin{bmatrix} 
       \bm{x}^\top \\ \bm{y}^\top 
       \end{bmatrix} &= \bm{s}^\top, \\ 
\begin{bmatrix}
I_{n_{2}} & \bm{0} & \mathcal{I}M(\bm{h})^\top \\ \bm{0} & I_{n_{2}} & \mathcal{I}M(\bm{s})^\top 
\end{bmatrix} 
\begin{bmatrix}
\bm{R}_1 \\ \bm{E} \\ \bm{R}_{2} 
\end{bmatrix}
&= \begin{bmatrix} 
\bm{U} \\ \bm{V} - Fold(\bm{m}G)
\end{bmatrix}.
\end{align*} $\hfill \blacksquare$    
\end{proof}

\subsection{RQC from the EGK codes with non-homogeneous errors and unstructured rank}
\textbf{Protocol Description}:
\label{rrr}
Before presenting our new scheme, we still need to the following notations. The notaion $ \mathcal{S}_{(\omega_{1},\omega_{2})}^{a \times (b,c,d)}$ and $Fold()$ procedure, $Unfold()$ procedure refer to the procedure introduced in Section \ref{rr}.

Our second new encryption scheme with multiple syndromes and uses the following two codes:

$\star$ An $[n,k]_{q^m}$-EGK code $G \in \mathbb{F}_{q^{m}}^{k_{1}k_{2} \times n_{1}n_{2}}$ generated by $\bm{g}_{1} \in \mathbb{F}_{q^m}^{n_{1}}$ with weight $t_{1}$ and $\bm{g}_{2} \in \mathbb{F}_{q^m}^{n_{2}}$ with weight $t_2$. This code can correct errors of weight $r \leq \left \lfloor \frac{t_{2}-k_{2}}{2} \right \rfloor$ for the case of $k_{1}=t_{1},t_{2}=m<t_{1}t_{2}$, and for the case of $k_{1}\leq t_{1},k_{2}\leq t_{2},t_{1}t_{2}\leq m$, $r \leq  \frac{(t_{1}-k_{1}+1)(t_{2}-k_{2}+1)-1}{2}$ in a deterministic way.

$\star$  A random $[2z,z]_{\mathbb{F}_{q^m}}$-ideal code, of parity matrix $[I_{z} ~ H]$.

    For a matrix $A \in \mathbb{F}_{q^m}^{c \times d}$ (with rows $A_{1},\ldots,A_{c}$) and a matrix $B \in \mathbb{F}_{q^m}^{d \times b}$, we define the product $A \cdot B \in \mathbb{F}_{q^m}^{c \times b}$ as the matrix whose $i$-th row is $A_{i} \cdot B$, where the row vector $A_{i}$ multiplies the matrix $B$ using the vector-matrix product defined in Section \ref{rr}. Next, our proposed RQC.EGK-Multi-NH is presented in Figure \ref{fig:11}.

\begin{figure}[h]
\begin{minipage}{\linewidth}
\begin{center}
    \fbox{ 
        \begin{minipage}{0.9\textwidth} 
            \underline{RQC.Setup($1^{\lambda}$):}\\
            \hspace{2em} $\bullet$ Generates and outputs the global parameters $param = (z,k,n_{1},n_{2},k_{1},k_{2},t_{1},t_{2},\omega_{\bm{x}},\omega_{\bm{y}},\omega_{1},\omega_{2},r)$.\\
            \underline{RQC.KGen($\lambda$):}\\
            \hspace{1cm} $\bullet$ Sample ${\bm{X},\bm{Y}}\stackrel{\text{\$}}{\leftarrow}\mathcal{S}_{\omega_{\bm{x}},\omega_{\bm{y}}}^{z \times (n_{1},n_{1})}$;\\
            \hspace{2em} $\bullet$ Sample $\bm{g}_{1} \stackrel{\text{\$}}{\leftarrow} \mathcal{S}_{t_{1}}^{n_{1}}(\mathbb{F}_{q^m})$, $\bm{g}_{2} \stackrel{\text{\$}}{\leftarrow} \mathcal{S}_{t_{2}}^{n_{2}}(\mathbb{F}_{q^m})$, $\bm{H} \stackrel{\text{\$}}{\leftarrow} \mathbb{F}_{q^m}^{z \times z}$; \\
            \hspace{2em} $\bullet$ Compute $\bm{S}=\bm{X}+\bm{HY}$;\\
            \hspace{2em} $\bullet$ Output a public key $pk = (\bm{g_{1}},\bm{g_{2}},\bm{H},\bm{S})$ and a private key $sk = (\bm{X},\bm{Y})$.\\
            \underline{RQC.Enc(pk, $\bm{m}$):}\\
             \hspace{2em} $\bullet$ Input the public key $pk$ and a plaintext $\bm{m} \in \mathbb{F}_{q^m}^{k}$;\\
             \hspace{2em} $\bullet$ Compute the generator matrix $\bm{G}_{1} \in \mathbb{F}_{q^m}^{k_{1} \times n_{1}}$ of the $[n_{1},k_{1}]_{q^m}$ Extended Gabidulin codes defined by $\bm{g}_{1}$;\\
              \hspace{2em} $\bullet$  Compute the generator matrix $\bm{G}_{2} \in \mathbb{F}_{q^m}^{k_{2} \times n_{2}}$ of the $[n_{2},k_{2}]_{q^m}$ Extended Gabidulin codes defined by $\bm{g}_{2}$;\\
               \hspace{2em} $\bullet$ Compute the generator matrix $\bm{G}= \bm{G}_{1} \otimes \bm{G}_{2} \in \mathbb{F}_{q^m}^{k \times n}$ of the $[n =n_{1}n_{2}, k= k_{1}k_{2}]_{q^m}$ Extended Gabidulin-Kronecker product codes defined by $\bm{g}_{1},\bm{g}_{2}$;\\
               \hspace{2em} $\bullet$ Use the randomness $\theta$ to generate $(\bm{R}_{1},\bm{E},\bm{R}_{2}) \stackrel{\text{\$}}{\leftarrow} \mathcal{S}_{(\omega_{1},\omega_{2})}^{n_{2} \times (z,n_{1},z)}$;\\
              \hspace{2em} $\bullet$ Compute $\bm{U}=\bm{R}_{1}+ \bm{R}_{2} \cdotp \bm{H} $ and $\bm{V} = Fold(\bm{m} \cdotp G)+ \bm{R}_{2} \cdotp \bm{S}+\bm{E}$;  \\
              \hspace{2em} $\bullet$  Output a ciphertext $\bm{c} = (\bm{U},\bm{V})$.  \\
            \underline{RQC.Dec(sk, $\bm{c}$):}\\
             \hspace{2em} $\bullet$ Input the private key $sk$ and the ciphertext $\bm{c}$;\\
             \hspace{2em} $\bullet$ Output the plaintext $\bm{m} := DecEKGCode\Big((\bm{g}_{1}\otimes\bm{g}_{2}),\bm{V}-\bm{Y} \cdotp \bm{U}\Big)$.
        \end{minipage} }
\end{center}
\end{minipage}
 \caption{Description of our proposal RQC.EGK-Multi-UR}
    \label{fig:11}
\end{figure}

$\textbf{Correctness}$:  The correctness of our encryption scheme clearly relies on the decoding capability of the Extended Gabidulin-Kronecker product code $\mathcal{C}$. Specifically, assume that code $\mathcal{C}$ can correctly decode $\bm{V}-\bm{Y \cdot U}$, then $$Dec\Big(sk,Enc(\bm{m},pk)\Big) = \bm{m}$$
holds. Meanwhile, for code $\mathcal{C}$ to correctly decode $\bm{V}-\bm{Y \cdot U}$, we have 
\begin{align*}
\nonumber
    \bm{V-Y} \cdot \bm{U} &= Fold(\bm{m}G) + \bm{S}\bm{R}_{2}+\bm{E}-\bm{Y}\cdot \bm{U}\\  
    &= Fold(\bm{m}G) + (\bm{X}+\bm{H}\cdot \bm{Y})\cdot \bm{R}_{2}+\bm{E}-\bm{Y}(\bm{R}_{1}+\bm{H} \cdot \bm{R}_{2}) \\ 
    &= Fold(\bm{m}G) + (\bm{X} \cdot \bm{R}_{2}-\bm{Y}\cdot \bm{R}_{1}+\bm{E}).
\end{align*}
Therefore, we have $Unfold(\bm{V}-Y \cdot \bm{U})= \bm{m}G+ Unfold(\bm{X} \cdot \bm{R}_{2}-\bm{Y} \cdot \bm{R}_{1}+ \bm{E}) \in \mathbb{F}_{q^m}^{n}$. The error term $Unfold(\bm{X} \cdot \bm{R}_{2}-\bm{Y} \cdot \bm{R}_{1}+\bm{E})$ must satisfy
\begin{equation*}
  wt_{R}(Unfold(\bm{X} \cdot \bm{R}_{2}-\bm{Y} \cdot \bm{R}_{1}+\bm{E})) \leq r.   
\end{equation*}
where $r \leq \left \lfloor \frac{t_{2}-k_{2}}{2} \right \rfloor$ for the case of $k_{1}=t_{1},t_{2}=m<t_{1}t_{2}$, and for the case of $k_{1}\leq t_{1},k_{2}\leq t_{2},t_{1}t_{2}\leq m$, $r \leq  \frac{(t_{1}-k_{1}+1)(t_{2}-k_{2}+1)-1}{2}$, and the message $\bm{m}$ can be correctly decoded. 

\begin{theorem}
     The scheme RQC.EGK-Multi-NH presented in Sec.\ref{rrr} is IND-CPA secure under the DRSL and DNHRSL problems. To be specific, for any probabilistic polynomial-time adversary $\mathcal{A}$, there exists a probabilistic polynomial-time adversary $\mathcal{B}$ such that the advantage of $\mathcal{A}$ against the IND-CPA experiment in our improved RQC is bounded by:
$$Adv_{EGK.RQC, \mathcal{A}}^{IND-CPA}(\lambda) \leq 2\Big(Adv_{\mathcal{B}}^{DRSL(m,2z,z,\omega_{x}+\omega_{y},n_{1})}(\lambda)+Adv_{\mathcal{B}}^{DNHRSL(m,\bm{\eta}_{3},z,\bm{\rho},n_{2})}(\lambda)\Big).$$
     
\end{theorem}
\begin{proof}
The proof of the RQC.EGK-Multi-UR scheme is similar to the proof from \cite{ref6} with a $RSL(m,2z,z,\omega_{x}+\omega_{y},n_{1})$ instance defined from a $[2z,z]$ code and a $NHRSL(m,\bm{\eta}_{3},z,\bm{\rho},n_{2})$ instance defined from an $[2z+n_{1},z]$ code, where $\bm{\eta}_{3}=(z,n_{1},z)$, $\bm{\rho}_{2}=(\omega_{1},\omega_{2})$. Two instances are defined as 

\begin{align*}
       \begin{bmatrix} 
       \bm{I}_{z} & \bm{H} 
       \end{bmatrix}
       \begin{bmatrix} 
       \bm{X} \\ \bm{Y}
       \end{bmatrix} &= \bm{S}, \\ 
[\bm{R}_{1},\bm{E},\bm{R}_{2}]
\begin{bmatrix}
\bm{I}_z &  \bm{0}\\ \bm{0}  & \bm{I}_{n_{1}}\\ \bm{H} & \bm{S} 
\end{bmatrix}
&= \begin{bmatrix} 
\bm{U} & \bm{V} - Fold(\bm{m}G)
\end{bmatrix}.
\end{align*} $\hfill \blacksquare$    
\end{proof}

\section{Security and Parameters of Our Proposed Schemes}
\label{sec777}
\subsection{Security analysis}
\subsubsection{Combinatorial attacks on the Ideal Blockwise Rank Syndrome Decoding (IBRSD) problem}
When $l=1$, the BRSD problem is equivalent to the RSD problem. Over the years, the RSD problem has been widely studied. Combinatorial attacks are a mainstream class of attack methods targeting the RSD problem. Their core idea is to combine combinatorial search with linear algebraic techniques to recover the support set of the error vector in polynomial time. The key steps of combinatorial attacks can be summarized as an iterative process of support set guessing and linear system solving.  \textbf{Support set guessing}-The attacker first assumes that the support set of the error vector satisfies certain structural conditions and gradually narrows down the search space through probabilistic methods. For example, initial attack methods randomly sample candidate support sets and screen valid solutions by verifying syndrome consistency. \textbf{Construction of linear systems}-Under the premise that the support set assumption holds, the original problem is transformed into a system of linear equations. Specifically, the non-zero elements of the error vector are treated as variables, and syndrome equations provide linear constraints, thereby forming an underdetermined or overdetermined linear system. \textbf{Solution via Gaussian elimination}-The aforementioned system is solved using Gaussian elimination or other linear algebra algorithms. If the support set assumption is correct, the error vector can be uniquely determined. If the assumption is incorrect, backtracking is performed and the guessing strategy is adjusted.

The advantage of such attacks is that they avoid directly dealing with NP-hard problems, reducing the complexity to the polynomial level through heuristic search and linear algebra tools, but at the cost of introducing a certain failure probability. The first combinatorial attack paper \cite{ref33} proposed a systematic attack framework targeting the RSD problem. This method works by randomly guessing the row support set of the error matrix and solving a linear system. In 2002, an improved method was proposed in \cite{ref15}. This attack transforms the quadratic multivariate problem obtained from the RD problem into a linear system by guessing the entries of the support matrix and the coefficient matrix. Later in 2016, Reference \cite{ref16} refined the results of the current combinatorial attack in a more elaborate manner. The key to their method lies in more ingeniously guessing the support of the error and solving the linear system. Two years later, Aragon et al. further improved the above results in \cite{ref17}.

References \cite{song2025interleaved} presents several forms for solving the BRD problem, which can be summarized as follows: 

$\star$ Solving the BRD problem consists in finding an $\bm{u} \in \mathbb{F}_{q^m}^{k+1}$ such that
\begin{equation}
\label{7}
    \bm{u} \cdot \begin{pmatrix} \bm{y} \\G\end{pmatrix}
    = \bm{e}, ~~where~ G_{y}= \begin{pmatrix} \bm{y} \\G\end{pmatrix}.
\end{equation}

$\star$ Solving the BRD problem consists in finding an blockwise error $\bm{e}$ of weight $r$ such that 
\begin{equation}
\label{8}
    \bm{e}H_{y}^{T} = \bm{0}_{n-k-1},
\end{equation}

where $H_{y}$ is the parity-check matrix of $G_{y}$.

$\star$ The error vector $\bm{e}$ can be expressed as follows, 
\begin{equation}
\label{9}
    \bm{e}= \alpha \bm{SC} = \alpha  
\begin{pmatrix} 
\bm{I}_r \\  \hline
\bm{0}_{(m-r) \times 1} & \vert & \bm{S' }
\end{pmatrix} 
\begin{pmatrix}
\begin{array}{c|c}
 1 &  \bm{C'} \\  
\bm{0}_{(r-1) \times 1} &  
\end{array}  
\end{pmatrix}
\end{equation}

$\star$ The error vector $\bm{e}$ can be expressed as follows, 
\begin{equation}
\label{10}
    \bm{e}= \alpha \bm{SC} = \alpha  
\begin{pmatrix}
\begin{array}{c|c}
 1 &  \bm{C'} \\  
\bm{0}_{(r-1) \times 1} &  
\end{array}  
\end{pmatrix}
\begin{pmatrix}
    \bm{Q}_{1}\bm{C}_{1} & \bm{0} &\cdots & \bm{0}\\
    \bm{0} & \bm{Q}_{2}\bm{C}_{2} & \cdots & \bm{0}\\
    \vdots & \vdots & \ddots  & \vdots\\
    \bm{0} & \bm{0} & \cdots & \bm{Q}_{l}C_{l}
\end{pmatrix}
\end{equation}

where $S$ is the support matrix and $C$ is the coefficient matrix.

For each of the above formulations (\ref{7}),(\ref{8}),(\ref{9}),(\ref{10}), corresponding solution methods are provided, namely AGHT attack, PRR attack and OJ attack, they are summarized in Table \ref{tab:22}.

\begin{table}[htbp]
\small
\centering
\caption{Some known combinatorial attacks on the BRD problem.}
\label{tab:22}       
\begin{tabular}{lll}
\hline\noalign{\smallskip}
Attacks & Conditions & Complexity  \\
\noalign{\smallskip}\hline\noalign{\smallskip}
AGHT & $m \leq n$ & $\mathcal{O}\bigg(\Big(n-k-1)m\Big)^{\omega}q^{r\left\lceil\frac{(k+1)m}{n}\right\rceil - m}\bigg)$ \\
 & $m > n $ & $\mathcal{O}\Big((n-k-1)^{\omega}m^{2}q^{r(k+1)}\Big)$ \\
PRR & $r_{i} \leq t_{i} \leq m $ for $1 \leq i \leq l$, $\sum_{i=1}^{l}t_{i} \leq m$, $\sum_{i=1}^{l}n_{i}t_{i} \leq m(n-k-1)$   & $\mathcal{O}\bigg(\Big(m(n-k-1)\Big)^{\omega}q^{\sum_{i=1}^{l}r_{i}(m-t_{i})-m}\bigg)$ \\
 &  $r_{i} \leq t_{i}^{'} \leq n_{i}$ for $1 \leq i \leq l$, $\sum_{i=1}^{l}t_{i}^{'} \leq n-k-1$  & $\mathcal{O}\bigg((n-k-1)^{\omega}m^{2}q^{\sum_{i=1}^{l}r_{i}(n_{i}-t_{i}^{'})}\bigg)$ \\
OJ & basis enumeration   & $\mathcal{O}\bigg((kr+r)^{\omega}q^{(m-r)(r-1)}\bigg)$,   \\
 &  $n_{1}-1 \leq k < n_{1}+r_{2}-1$  &  $\mathcal{O}\bigg(\Big(m(r-1)+(n_{1}-r_{1})\Big)^{\omega}q^{(r_{1}-1)(n_{1}-r_{1})+\gamma}\bigg)$, \\
& $r_{1}-1 \leq k \leq n_{1}-1$ & $\mathcal{O}\bigg(\Big(m(r-1)+(k+1-r_{1})\Big)^{\omega}q^{(r_{1}-1)(k+1-r_{1})+\gamma}\bigg)$,\\
 & $1 \leq k \leq r_{1}-1$   & $\mathcal{O}\bigg(\Big(m(r-1)\Big)^{\omega}q^{\gamma}\bigg)$, \\
 & & where $\gamma = max\{r_{i}: 2 \leq i \leq l\}$ and $r = \sum_{i=1}^{l}r_{i}$\\
\noalign{\smallskip}\hline
\end{tabular}
\end{table}
\begin{remark}
  This attack is a general attack against the BRSD problem, and no better improvement methods exist when the ideal structure of the code is adopted.  
\end{remark}
\subsubsection{Algebraic attacks on the Ideal Blockwise Rank Syndrome Decoding (IBRSD) problem}
The algebraic attack methods for solving the BRSD problem originate from those for solving the RSD problem. The core logic of algebraic attacks lies in constructing a system of constraint equations based on annihilator polynomials, transforming the goal of finding a low-rank error vector in the RD problem into a task of solving an algebraic system. Specifically, by leveraging the rank constraint properties of the error vector, attackers construct a set of polynomials that satisfy the condition of yielding zero when acting on the error matrix. They then simplify the solution process through two key techniques. The first technique is linearization technique. This transforms nonlinear polynomial equations into linear systems of equations, lowering the threshold for initial solution. It is particularly effective for handling low-degree constraint relations. The second technique is 
Gr$\ddot{o}$bner basis method. Dubbed the Gaussian elimination for solving polynomial systems, it converts complex equation systems into a triangulated canonical form using Buchberger's Algorithm. This enables the stepwise elimination of variables and efficient determination of solutions.

Early research on algebraic attacks was represented by \cite{ref16}. Although its modeling approach achieved the algebraic transformation of the RD problem, it suffered from a long-standing efficiency bottleneck. For a considerable period of time, such methods were generally regarded as inferior to combinatorial attacks—especially when the size of the finite field $\mathbb{F}_q$ is small. The number of variables and the degree of equations in the polynomial system grow sharply, leading to an exponential increase in the time complexity of Gr$\ddot{o}$bner Basis computations, which makes it difficult to adapt to practical cryptographic parameters. This limitation confined algebraic attacks to the theoretical level for a long time and prevented them from posing a substantive threat to rank-based cryptosystems. The study in \cite{ref18} broke the performance bottleneck of algebraic attacks. Its core innovation lies in the design of a targeted MM modeling method by leveraging $\mathbb{F}_{q^m}$-linearity. Unlike early universal modeling approaches, MM modeling fully explores the structural characteristics of the RD problem. It constructs sparse polynomial systems by extracting the maximal minors of the error matrix, significantly reducing the number of variables and the complexity of equations. This specialized design enabled algebraic attacks to demonstrate the potential to outperform combinatorial attacks in certain parameter scenarios for the first time, and it has even been successfully applied to the cryptanalysis of Rollo and RQC—candidates in the second round of NIST post-quantum standardization. Building on MM modeling, \cite{ref19} further proposed SM modeling, extending the applicability of algebraic attacks from the RD problem to the broader class of MinRank problems. As a universal modeling framework, SM modeling constructs constraint equations by characterizing the support space of the error vector, without relying on the specific structural properties of individual problems. Thus, it can be migrated to other fields dependent on the MinRank hard problem, such as multivariate cryptography—for instance, SM modeling has shown remarkable effectiveness in the analysis of the Rainbow signature scheme. Its universality fills the adaptation gap of specialized modeling in cross-problem scenarios and improves the methodological system of algebraic attacks. To balance the efficiency of specialized modeling and the adaptability of universal modeling, subsequent studies (e.g., \cite{ref47}) integrated MM and SM modeling, proposing the SM-$\mathbb{F}_{q^m}^+$ hybrid modeling method. Through variable reduction techniques over the $\mathbb{F}_{q^m}$ field, this method inherits the sparsity advantage of MM modeling while retaining SM modeling's ability to characterize complex rank constraints. It effectively addresses the shortcomings of single modeling approaches in terms of parameter adaptability or solution efficiency.

The following Table \ref{tab:222} presents the complexity of existing algebraic attacks on the BRD problem.
\begin{table}[htbp]
\small
\centering
\caption{Some known algebraic attacks on the BRD problem.}
\label{tab:222}       
\begin{tabular}{lll}
\hline\noalign{\smallskip}
Attacks & Conditions & Complexity  \\
\noalign{\smallskip}\hline\noalign{\smallskip}
AP & linearization & $\mathcal{O}\bigg(min\Big\{(r_{v}k)^{\omega}q^{r_{v}\left \lceil \frac{(k+1)(r_{v}+1)-(n_{v}+1)}{r_{v}} \right \rceil}:v \in \{1,\ldots,l\}\Big\}\bigg)$ \\
 & Gr$\ddot{o}$bner basis & $\mathcal{O}\bigg(min\Big\{n_{v} \binom{r_{v}+k+d_{reg}^{(v)}-1}{d_{reg}^{(v)}}^{\omega}\Big\}:v \in \{1,\ldots,l\}\bigg)$ \\
MM  & $m\binom{n-k-1}{r} \ge \prod_{i=1}^{l} \binom{n_{i}}{r_{i}}-1$   & $\mathcal{O}\bigg(m\binom{n-p-k-1}{r}\Big(\binom{n_{l}-p}{r_{l}}\prod_{i=1}^{l-1} \binom{n_{i}}{r_{i}}\Big)^{\omega-1}\bigg)$, \\
& & where $p = max\Big\{i \Big| m\binom{n-i-k-1}{r} \ge \binom{n_{l}-i}{r_{l}}\prod_{i=1}^{l-1} \binom{n_{i}}{r_{i}}-1 \Big\}$\\
 &  $m\binom{n-k-1}{r} < \prod_{i=1}^{l} \binom{n_{i}}{r_{i}}-1$  & $\mathcal{O}\bigg(q^{\sum_{i=1}^{l}a_{i}r_{i}}m\binom{n-k-1}{r}\Big(\prod_{i=1}^{l}\binom{n_{i}-a_{i}}{r_{i}}\Big)^{\omega-1}\bigg)$, \\
 & & where $(a_{1},a_{2},\ldots,a_{l})$ satisfies $m \binom{n-k-1}{r} \ge \prod_{i=1}^{l} \binom{n_{i}-a_{i}}{r_{i}}-1$\\
\noalign{\smallskip}\hline
\end{tabular}
\end{table}

To further reduce the attack complexity, the References \cite{song2025interleaved} proposed the BP method for the BRD problem, achieving a lower complexity. The relevant results are as follows.
\begin{theorem}
    Let $\bm{a} = (a_{1},a_{2},\ldots,a_{l})$ and $\bm{p}=(p_{1},p_{2},\ldots,p_{l})$ be two integers vectors and $p = \sum_{i=1}^{l}p_{i}$. The MM model for solving the BRD problem, when using the BP method, has a complexity of: 
    \begin{equation}
        \mathcal{O}\bigg(q^{\sum_{i=1}^{l}a_{i}r_{i}}m\binom{n-p-k-1}{r}\Big(\prod_{i=1}^{l}\binom{n_{i}-p_{i}-a_{i}}{r_{i}}\Big)^{\omega-1}\bigg).
        \nonumber
    \end{equation}
where $\bm{a},\bm{p}$ and $p$ such that $m \binom{n-p-k-1}{r} \ge \prod_{i=1}^{l}\binom{n_{i}-p_{i}-a_{i}}{r_{i}}$ holds and the complexity is minimal.
\end{theorem} $\hfill\blacksquare$
\begin{remark}
    For the BRSD problem, when $l=1$, it is equivalent to the general RSD problem.
\end{remark}

\subsubsection{Attack on the NHRSD problem}
Reference \cite{ref7} presented the first detailed analysis on solving the NHRSD problem. By leveraging the non-homogeneous structure of errors, two types of attacks were proposed, namely combinatorial attacks and algebraic attacks. 

The core idea of combinatorial attacks is to leverage the non-homogeneous structure of errors (error vector partitioned as $\bm{e} = (\bm{e}_{2},\bm{e}_{1},\bm{e}_{3})$, with support spaces $\bm{S}_{1} = Supp(\bm{e}_{1},\bm{e}_{3})$ and $\bm{S}_{2} = Supp(\bm{e}_{2})$, and transform the problem into solving a linear system via subspace guessing. The specific method consists of the following four steps. First, guess two subspaces $V$ of dimension $r \ge \omega_{1}$  (satisfying $S_{1} \subset V$) and $Z$ of dimension $\rho \in \{\omega_{2},\ldots,m-r\}$ (satisfying $S_{2} \subset V \oplus Z)$. Second, draw on the techniques of \cite{ref17}, introduce $\alpha \in \mathbb{F}_{q^m}^{*}$ to improve the success probability (requiring $\alpha S_{1} \subset V$ and $\alpha S_{2} \subset V \oplus Z$). Third, construct a linear system using parity-check equations, express $\bm{e}_{1},\bm{e}_{3}$ as variables in the basis of $V$ and $\bm{e}_{2}$ as variables in the basis of $V \oplus Z$, and ensure a unique solution through the constraint (number of variables $\leq$ number of equations). Fourth, optimize the parameter pair $(r,\rho)$ and maximize the objective function to minimize the exponential complexity.

The core idea of algebraic attacks is based on the MaxMinors linear system (system of full-rank minor equations), simplify variables and equations using the special block structure of the error matrix, and solve the linear system via field projection. The specific method consists of the following four steps. First, utilize the block structure of the error matrix $C$ (containing zero subblocks) to identify and remove minor variables that are identically zero to reduce system complexity.
Second, partition the MaxMinors equations into three categories, $P_{lost}$ (identically zero, discarded), $P_{indep}$ (linearly independent, retained), and $P_{rest}$ (independent of $P_{indep}$, retained). Third, project the remaining equations onto the base field $\mathbb{F}_{q}$ to obtain an expanded system ($P_{indep,\mathbb{F}_{q}}$ and $P_{rest},\mathbb{F}_{q}$) with $m$ times the number of original equations. Fourth, first simplify $P_{rest,\mathbb{F}_{q}}$, substitute variables using its row echelon basis, then solve the simplified $P^{'}_{indep,\mathbb{F}_{q}}$, if there are insufficient equations, supplement constraints by fixing columns of matrix $C$ (optimize the exponential factor using the error structure).

The complexity of solving the NHRSD problem is summarized in Table \ref{tab:22223}.
\begin{table}[htbp]
\small
\centering
\caption{Some known attacks on the NHRSD problem.}
\label{tab:22223}       
\begin{tabular}{lll}
\hline\noalign{\smallskip}
Attacks & Complexity  \\
\noalign{\smallskip}\hline\noalign{\smallskip}
Combinatorial attacks  & $\mathcal{O}\Big(q^{(\omega_{1}+\omega_{2})(m-r)-\omega_{2}\rho-m}\Big)$, \\
  & where $\omega_{1} \leq r, \omega_{2} \leq \rho, r+\rho \leq m-1$. \\
Algebraic attacks    & $\mathcal{O}\bigg(q^{a \omega_{1}}\mathcal{N}_{\mathbb{F}_{q}}\Big(\binom{2n+n_{1}-a}{\omega_{1}+\omega_{2}}-M_{a}-v_{\mathbb{F}_{q}}\Big)^{\omega-1}\bigg)$, \\
 & where $\omega$ is a linear algebra constant, and $\mathcal{N}_{\mathbb{F}_{q}} = m\sum_{i=\omega_{2}}^{\omega_{1}+\omega_{2}} \binom{n_{1}-1}{i}\binom{n}{\omega_{1}+\omega_{2}-i}$, \\
 & $v_{\mathbb{F}_{q}} = m \binom{n_{1}-1}{\omega_{2}-1} \binom{n-1}{\omega_{1}}$, $M_{a} = \sum_{i=0}^{\omega_{2}-1} \binom{n_{1}}{i}\binom{2n-a}{\omega_{1}+\omega_{2}-i}$,\\
& $a \ge 0$ is the smallest integer such that $\mathcal{N}_{\mathbb{F}_{q}} \ge \binom{2n+n_{1}-a}{\omega_{1}+\omega_{2}}-M_{a}-v_{\mathbb{F}_{q}}-1$.\\
\noalign{\smallskip}\hline
\end{tabular}
\end{table}

To further reduce the attack complexity, the References \cite{song2025interleaved} proposed the BP method for the NHRD problem, achieving a lower complexity. The relevant results are as follows.
\begin{theorem}
    Let $\bm{a} = (0,0,a)$ and $\bm{p}=(p_{1},p_{2},p_{3})$ be two integers vectors and $p = \sum_{i=1}^{3}p_{i}$. The MM model for solving the NHRD problem, when using the BP method, has a complexity of: 
    \begin{equation}
        \mathcal{O}\bigg(q^{ar_{1}}m\binom{n-p-k-1}{r}\Big(U(n-p-a,k,r,\bm{n}-\bm{p}-\bm{a},\bm{r})\Big)^{\omega-1}\bigg),
        \nonumber
    \end{equation}
where $\bm{a},\bm{p}$ and $p$ such that $m \binom{n-p-k-1}{r} \ge U(n-p-a,k,r,\bm{n}-\bm{p}-\bm{a},\bm{r})$ holds and the complexity is minimal, and $\bm{n} = (n_{1},n_{2},n_{3})$, $\bm{r} = (r_{1},r_{2})$, $n=\sum_{i=1}^{3}n_{i}$, $r = \sum_{i=1}^{2}r_{2}$,
\begin{equation}
\nonumber
U(n,k,r,\bm{n},\bm{r})=\left\{
\begin{aligned}
\binom{n}{r}-\sum_{i=0}^{r_{2}-1}\binom{n_{2}}{i}\binom{n_{1}+n_{3}}{r-i},& ~~if~ n_{2}> r_{2}, n_{1}+n_{3}\ge r_{1}  \\
\binom{n_{2}}{r_{2}}\binom{n_{1}+n_{3}}{r_{1}}, & ~~ if ~n_{2}=r_{2}, n_{1}+n_{3}\ge r_{1}\\
\sum_{i=0}^{n_{1}+n_{3}}\binom{n_{2}}{r-i}\binom{n_{1}+n_{3}}{i}, & ~~ if~ n_{2}>r_{2}, n_{1}+n_{3} \leq r_{1}
\end{aligned}
\right. 
\end{equation}  $\hfill\blacksquare$
\end{theorem} 

\subsubsection{Attack on the RSL problem}
The RSL problem is a core hard problem in rank-metric cryptography. Its goal is to recover the support space of the error vector (i.e., the linear subspace spanned by the error vector) rather than the error vector itself. Early research focused on the RSL problem with homogeneous errors, where the main attack method was naive subspace guessing, which suffered from extremely high complexity. Later, with the application of the non-homogeneous error model in cryptographic schemes, attacks were extended to NHRSL. By leveraging the non-homogeneous structure of errors (constant term + homogeneous component), the guessing strategy was optimized to reduce complexity. Currently, there are two main approaches for solving the RSL problem, namely combinatorial attacks and algebraic attacks. 

The core idea of combinatorial attacks is based on dimensional constraints of the support space. Through subspace sampling and syndrome verification, the range of candidate support spaces is gradually narrowed down to ultimately locate the target subspace. The specific method mainly consists of the following three steps. First, utilize the support space structure of non-homogeneous errors ($S=S_{1} \oplus S_{2}$, where $S_{1}$ is the support space of the homogeneous component and $S_{2}$ is the support space of the non-homogeneous component) to guess $S_{1}$ and $S_{2}$ in phases. Second, introduce the linear transformation technique of $\alpha \in \mathbb{F}_{q^m}^{*}$ to expand the sampling coverage and improve the attack success rate. Third, screen invalid candidate support spaces through syndrome equations to reduce redundant computations. The complexity of the combinatorial attack proposed in \cite{ref7} is $\mathcal{O}\Big(q^{r(m- \left \lfloor \frac{m(n-k)-N}{n-a} \right \rfloor}\Big)$, where $a:=\left \lfloor\frac{N}{r}\right \rfloor$, and $N < kr$.

The core idea of the algebraic Attack is to transform the RSL problem into polynomial system solving. It leverages the algebraic properties of the support space (such as linear constraints of subspaces and matrix rank conditions) to construct equations, which are then solved using linearization or Gröbner basis methods. Specific methods include linearized equation construction and the MaxMinors extension attack. Among them, linearized equation construction consists of two steps. Step 1: Let the basis of the support space $S$ be $\{\bm{b}_{1},\ldots,\bm{b}_{\omega}\}$, and the error vector $\bm{e} = \sum_{i=1}^{\omega}x_{i}\bm{b}_{i}$ ($x_{i} \in \mathbb{F}_{q^m}$). Step 2: Combine the syndrome equation $H \bm{e}^{T}= \bm{s}$, treat the components of the basis vectors as variables, and construct a system of linearized polynomial equations. The MaxMinors extension attack also includes two steps. Step 1: Draw on the idea of the MaxMinors system for NHRSD to construct a system of minor equations related to the support space. Step 2: Utilize the zero minor structure of non-homogeneous errors to remove redundant variables and reduce the size of the polynomial system. The complexity of the algebraic attack proposed in \cite{ref7} for `$\delta = 0$' is:
\begin{equation}
\nonumber
    \mathcal{O}\bigg(min\Big(2^{r\alpha_{R}+\alpha_{\lambda}}m \mathcal{N}_{\leq b}^{\mathbb{F}_{2}}(\mathcal{M}_{\leq b}^{\mathbb{F}_{2}})^{\omega-1},2^{r\alpha_{R}+\alpha_{\lambda}}(N^{'}-\alpha_{\lambda})\binom{k-a+1+r}{r}\Big)(\mathcal{M}_{\leq b}^{\mathbb{F}_{2}})^{2})\bigg)
\end{equation}
where $a$ is the unique integer such that $ar < N \leq (a+1)r$, $N^{'} = ar +1$, $1 \leq b \leq r+1$, $0 \leq \alpha_{R} < n-a-r$, $0 \leq \alpha_{\lambda} < N^{'} -b$, 
\begin{equation}
\nonumber
    \mathcal{M}_{\leq b}^{\mathbb{F}_{2}} = \sum_{i=1}^{b}\binom{n-a-\alpha_{R}}{r}\binom{N^{'}-\alpha_{\lambda}}{i},
\end{equation}
\begin{equation}
\nonumber
    \mathcal{N}_{\leq b}^{\mathbb{F}_{2}} = \sum_{i=1}^{b}\sum_{d=1}^{i}\sum_{j=1}^{n-k}\binom{j-1}{d-1}\binom{n-k-j}{r-d+1}\binom{N^{'}-\alpha_{\lambda}-j}{i-d}.
\end{equation}
and $m \mathcal{N}_{\leq b}^{\mathbb{F}_{2}} \ge \mathcal{M}_{\leq b}^{\mathbb{F}_{2}}-1$, the values of $b,\alpha_{R}$, $\alpha_{\lambda}$ are chosen to minimize the complexity.

The complexity of the algebraic attack proposed in \cite{ref7} for `$\delta > 0$' is:
\begin{equation}
\nonumber
    \mathcal{O}\bigg(min\Big(2^{r\alpha_{R}+\alpha_{\lambda}}m \mathcal{N}_{\leq b}^{\mathbb{F}_{2}}(\mathcal{M}_{\leq b}^{\mathbb{F}_{2}})^{\omega-1},2^{r\alpha_{R}+\alpha_{\lambda}}(N^{'}-\alpha_{\lambda})\binom{k-a+1+r}{r}\Big)(\mathcal{M}_{\leq b}^{\mathbb{F}_{2}})^{2})\bigg)
\end{equation}
where $\delta$ is a positive integer such that $N \ge \delta(n-r+\delta)$, $a$ is the greatest integer such that $N \ge \delta(n-r+\delta)+a(r-\delta)$, $N^{'} = \delta(n-r'+\delta)+a(r-\delta)$, $1 \leq b \leq r+1$, $0 \leq \alpha_{R} < n-a-r$, $0 \leq \alpha_{\lambda} < N^{'} -b$, 
\begin{equation}
\nonumber
    \mathcal{M}_{\leq b}^{\mathbb{F}_{2}} = \sum_{i=1}^{b}\binom{n-a-\alpha_{R}}{r-\delta}\binom{N^{'}-\alpha_{\lambda}}{i},
\end{equation}
\begin{equation}
\nonumber
    \mathcal{N}_{\leq b}^{\mathbb{F}_{2}} = \sum_{i=1}^{b}\sum_{d=1}^{i}\sum_{j=1}^{n-k}\binom{j-1}{d-1}\binom{n-k-j}{r-\delta-d+1}\binom{N^{'}-\alpha_{\lambda}-j}{i-d}.
\end{equation}
and $m \mathcal{N}_{\leq b}^{\mathbb{F}_{2}} \ge \mathcal{M}_{\leq b}^{\mathbb{F}_{2}}-1$, the values of $b,\alpha_{R}$, $\alpha_{\lambda}$ are chosen to minimize the complexity.

\subsubsection{Attack on the NHRSL problem}
 
Reference \cite{ref7} presented the first complexity result for the NHRSL problem by adapting the process of solving the RSL problem via combinatorial attacks to the non-homogeneous error under the condition $n<3z$, which can be shown as follows:
\begin{equation}
    \mathcal{O}\Big(q^{\omega_{2}(m-r)-(\omega_{2}-\omega_{1})\rho}\Big),
    \nonumber
\end{equation}
where $r$ and $\rho$ are integers chosen to maximize the quantity $(\omega_{2}r+(\omega _{2}-\omega_{1})\rho$ under the following constraints: $$N_{1}, N_{2}, r,\rho \in \mathbb{N}, N_{1}+N_{2} = n_{2}, \omega_{1} \leq r, \omega_{2}-\omega_{1} \leq \rho, r
+\rho \leq m-1$$ $$a = \left \lfloor \frac{N_{1}}{\omega_{1}} \right \rfloor \leq n-2z, b = \left \lfloor \frac{N_{2}}{\omega_{2}} \right \rfloor \leq 2z, m(n-z) \ge (n-2z-b)(r+\rho)+(2z-a)r+n_{2}.$$

\subsection{Parameters of Our Proposed Schemes}
\label{sec111}
In this subsection, we consider some practical attacks mentioned above, and present parameters corresponding to the 128, 192, 256 bits security level in Table \ref{tab0},\ref{tab1},\ref{tab2}. For the proposed RQC.EGK-BWE scheme, the public key size is ($\left \lceil \frac{mn}{8} \right \rceil +40$) bytes, and the ciphertext size is 2 $\left \lceil \frac{mn}{8} \right \rceil$ bytes. For the proposed RQC.EGK-Multi-NH scheme, the public key size is $\left \lceil \frac{m(t_{1}+t_{2}+2n_{2})+t_{1}n_{1}+t_{2}n_{2}}{8} \right \rceil$ bytes, and the ciphertext size is $2\left \lceil \frac{mn}{8} \right \rceil$ bytes. For the proposed RQC.EGK-Multi-UR scheme, the public key size is $\left \lceil \frac{m(t_{1}+t_{2}+z^{2}+n_{1}z)+t_{1}n_{1}+t_{2}n_{2}}{8} \right \rceil$ bytes, and the ciphertext size is $\left \lceil \frac{m(zn_{2}+n)}{8} \right \rceil$ bytes. Recall that the public key is contained in $(\bm{g}_{1}, \bm{g}_{2}, \bm{h},\bm{s})$ or $(\bm{g}_{1}, \bm{g}_{2}, \bm{H},\bm{S})$, and the ciphertext is contained in the pair $(\bm{u}, \bm{v})$ or $(\bm{U},\bm{V})$. The secret key is always generated from a 40 bytes seed, so the size of the private key is irrelevant.

To comprehensively evaluate the performance of our three proposed EGK-based RQC schemes, we present their public parameter sizes (public key and ciphertext) for the 128-bit, 192-bit, and 256-bit security levels in Table \ref{tab3}. A comparison is made with other prominent code-based cryptosystems, including: Classic McEliece \cite{ref22}, HQC \cite{ref20}, BIKE \cite{ref21}, LOI17 \cite{ref12}, Classic RQC \cite{song2025interleaved}, LT19 \cite{ref35}, Multi-RQC-AG \cite{ref7}, NH-Multi-RQC-AG \cite{ref7}, Multi-UR-AG \cite{ref7}, NH-Multi-UR-AG \cite{ref7}, BW-RQC \cite{ref10}. In the comparison, ``pk size" denotes public key size (in bytes), ``ct size" denotes ciphertext size (in bytes), and ``Total" denotes the sum of public key and ciphertext sizes (in bytes). All parameters are set based on the current best known attack complexities (including combinatorial and algebraic attacks) to ensure the target security level is achieved.

 At the 128-bit security level, the public key size of our RQC.EGK-BW schemes and RQC.EGK-Multi-NH schemes are approximately 3949 bytes and 3679 bytes, respectively. These are smaller than those of LOI17 (34560 bytes) and Multi-UR-AG (4114 bytes), and dramatically smaller than Classic McEliece (261120 bytes). Although larger than Classic RQC (860 bytes) and HQC (2249 bytes), they offer the advantage of zero decryption failure probability. Notably, the public key of RQC.EGK-Multi-UR is only 2138 bytes, which is smaller than HQC, LOI17, and NH-Multi-UR-AG, and represents a reduction of approximately 48\% compared to Multi-UR-AG. Furthermore, The public keys of RQC.EGK-BWE and RQC.EGK-Multi-NH are 4\% and 11\% smaller than Multi-UR-AG, respectively, demonstrating the effectiveness of the Kronecker structure in compressing public keys.

The ciphertext size of a scheme often correlates with its public key structure. For instance, Classic McEliece has a very small ciphertext but an extremely large public key, while RQC-family schemes typically feature a small public key and a moderate ciphertext size. A key feature of all three EGK-based schemes is their zero decryption failure probability, a property lacking in many existing schemes (e.g., HQC, BIKE, BW-RQC, EG-RQC). This property enhances long-term security by resisting attacks based on decryption failures (e.g., reaction attacks). Moreover, compared to the fourth-round NIST candidates Classic McEliece, HQC, and BIKE, our schemes show a clear advantage in public key size, especially at medium to high security levels. Compared to other improved variants of the RQC family (e.g., Multi-UR-AG, NH-Multi-RQC-AG), the EGK structure further compresses the public key while maintaining security and achieving zero decryption failure.

In summary, the parameter comparison demonstrates that our three proposed EGK-based RQC variants are competitive in terms of public key size, total transmission overhead, and decryption reliability. In particular, the RQC.EGK-Multi-UR scheme achieves a public key size of only 2138 bytes at the 128-bit security level, representing a significant improvement over current similar schemes while maintaining zero decryption failure probability. This makes it suitable for deployment in bandwidth-constrained environments with high-security requirements.

 \begin{table}[!h]
\caption{Parameters and public key size of RQC.EGK-BWE (in bytes).}
\centering
\label{tab0}   
\begin{tabular}{llllllllllllllllllll}
\hline\noalign{\smallskip}
\multicolumn{16}{c}{$Size_{pk}$ = $\left \lceil \frac{mn}{8} \right \rceil + 40$ bytes, $size_{ct} = 2 \left \lceil \frac{mn}{8} \right \rceil$ bytes}\\
\hline\noalign{\smallskip}
Scheme  & $n_{1}$ & $k_{1}$ & $n_{2}$ & $k_{2}$ & $q$ & $m$ & $n$ & $k$ & $r$ & $t_{1}$ & $t_{2}$ & $\omega_{\bm{x}}$ & $\omega_{\bm{y}}$ & $\omega_{1}$ & $\omega_{2}$ & $\omega_{\bm{e}}$ & $pk$ size & $ct$ size & Security\\
\noalign{\smallskip}\hline\noalign{\smallskip}
RQC.EGK-BWE  & 10  & 3  & 59  & 5  &  2 & 53  &  590 & 15 & 21 & 3  & 53 & 3 & 3
& 3    & 3  &  3 & 3949  & 7818  &   128\\
RQC.EGK-BWE  & 10  & 3  &  83 & 7  & 2  & 79  & 830  & 21 & 36 & 3 & 79  & 4  & 4  & 4   & 4  & 4  & 8237  & 16394  &  192\\
RQC.EGK-BWE  & 10  & 3  & 113  & 3  & 2  & 113  & 1130  & 9 & 55 & 3 & 113  & 5 & 5  &  5  &  5 & 5  & 16002  & 31924 &   256\\
\noalign{\smallskip}\hline
\end{tabular}
\end{table}

\begin{table}[!h]
\caption{Parameters and public key size of RQC.EGK-Multi-NH (in bytes).}
\centering
\label{tab1}  
\begin{tabular}{lllllllllllllllllll}
\hline\noalign{\smallskip}
\multicolumn{16}{c}{$Size_{pk}$ = $\left \lceil \frac{m(t_{1}+t_{2}+2n_{2})+t_{1}n_{1}+t_{2}n_{2}}{8} \right \rceil$ bytes, $size_{ct} =  2\left \lceil \frac{mn}{8} \right \rceil$ bytes}\\  
\hline\noalign{\smallskip}
Scheme   & $n_{1}$ & $k_{1}$ & $n_{2}$ & $k_{2}$ & $q$ & $m$ & $n$ & $k$ & $r$ & $t_{1}$ & $t_{2}$ & $\omega_{\bm{x}}$ & $\omega_{\bm{y}}$ & $\omega_{1}$ & $\omega_{2}$  & $pk$ size & $ct$ size & Security\\
\noalign{\smallskip}\hline\noalign{\smallskip}
RQC.EGK-Multi-NH     & 6  & 3  & 86  & 3  & 2  & 85  & 516 & 9 & 28  & 3 & 85 & 4  &  4  & 3  & 4  & 3679  & 10966 &   128\\
RQC.EGK-Multi-NH     & 6  & 3  & 99  & 3  & 2  & 97  & 594 & 9 & 45 & 3  & 97  & 5  &  5  & 4  & 5  & 4816  & 14406  &  192\\
RQC.EGK-Multi-NH     & 11  & 4  & 116  & 4  & 2  &  116 & 1276 & 16 & 56 & 4  & 116 & 5  & 5   & 5  & 6  & 6792  & 37004 &   256\\
\noalign{\smallskip}\hline
\end{tabular}
\end{table}

\begin{table}[!h]
\caption{Parameters and public key size of RQC.EGK-Multi-UR (in bytes).}
\centering
\label{tab2}   
\begin{tabular}{llllllllllllllllllll}
\hline\noalign{\smallskip}
\multicolumn{16}{c}{$Size_{pk}$ = $\left \lceil \frac{m(t_{1}+t_{2}+z^{2}+n_{1}z)+t_{1}n_{1}+t_{2}n_{2}}{8} \right \rceil$ bytes, $size_{ct} =  \left \lceil \frac{m(zn_{2}+n)}{8} \right \rceil$ bytes}\\ 
\hline\noalign{\smallskip}
Scheme   & $z$ & $n_{1}$ & $k_{1}$ & $n_{2}$ & $k_{2}$ & $q$ & $m$ & $n$ & $k$ & $r$ & $t_{1}$ & $t_{2}$ & $\omega_{\bm{x}}$ & $\omega_{\bm{y}}$ & $\omega_{1}$ & $\omega_{2}$ & $pk$ size & $ct$ size & Security\\
\noalign{\smallskip}\hline\noalign{\smallskip}
RQC.EGK-Multi-UR   &3  &  6 &  3 & 86  & 3  & 2  & 85  & 516 & 9 &  22 & 3 & 85 & 3  & 3   &  3 & 4  & 2138  & 8224 &   128\\
RQC.EGK-Multi-UR   & 3 &  6 & 3  & 92  & 3  & 2  & 91  & 552 & 9 & 41 & 3  &  91 & 4  & 4   & 4  &  9 & 2426  & 9419  &  192\\
RQC.EGK-Multi-UR &  3  & 6  & 4  & 117  & 4  & 2  & 116  & 702 & 16 & 56  & 4  & 116 & 5  &  5  & 5  & 6  & 3831  & 15269 &   256\\
\noalign{\smallskip}\hline
\end{tabular}
\end{table}

\begin{table}[!h]
\caption{Comparison of related schemes (in bytes).}
\label{tab3}
\centering
\begin{tabular}{lllllc}
\hline\noalign{\smallskip}
Instance  & Security & $pk$ size & $ct$ size & total & Decryption failure \\
\noalign{\smallskip}\hline\noalign{\smallskip}
HQC  \cite{ref20}             & 128      & 2249 & 4497 & 6746   & YES                 \\
BIKE   \cite{ref21}          & 128      & 1541  & 1573 & 3114   & YES                \\
Classic McEliece \cite{ref22} & 128      & 261120 & 96 & 261216  & NO                 \\
LOI17 \cite{ref12}           & 128      & 34560  & 313 & 34873  & NO                 \\
Classic RQC \cite{ref10} & 128 & 860 &1704  &2564  & NO \\
\textbf{RQC.EGK-BWE-128} & \textbf{128} & \textbf{3949}  & \textbf{7818} & \textbf{11767} & \textbf{NO}                 \\ 
\textbf{RQC.EGK-Multi-NH-128}  & \textbf{128} & \textbf{3679}  & \textbf{10966}  & \textbf{14645}    & \textbf{NO}    \\
\textbf{RQC.EGK-Multi-UR-128}  & \textbf{128} & \textbf{2138} & \textbf{8224}  & \textbf{10362}  & \textbf{NO}  \\
LT19 \cite{ref35}  & 128 & 2353 & 2353 & 4706 & NO\\
Multi-RQC-AG-128 \cite{ref7} & 128 & 435 & 3943 & 4378 & YES \\
NH-Multi-RQC-AG-128 \cite{ref7} & 128 & 422 &2288 & 2710 & YES \\
Multi-UR-AG-128 \cite{ref7} & 128 & 4114 & 6912 & 11026 & YES \\
NH-Multi-UR-AG-128 \cite{ref7} & 128 & 2650  & 4472 & 7122 & YES \\
BW-RQC \cite{ref10} & 128 & 860   & 1704  & 2564  & YES \\
EG-RQC \cite{song2025interleaved} & 128 & 796   & 1512  & 2308  & YES \\
\hline\noalign{\smallskip}
HQC  \cite{ref20}            & 192      & 4522  & 9042 & 13564   & YES                 \\
BIKE  \cite{ref21}            & 192      & 3083  & 3115 & 6198   & YES                \\
Classic McEliece \cite{ref22} & 192      & 524160 & 156 & 524316  & NO                 \\
LOI17  \cite{ref12}          & 192      &  --  & -- & --  & NO                 \\
Classic RQC \cite{ref10} & 192 & 1834 & 3652 & 5486 & NO \\
\textbf{RQC.EGK-BWE-192} & \textbf{192}   & \textbf{8237}  & \textbf{16394}  & \textbf{24631} & \textbf{NO}  \\ 
\textbf{RQC.EGK-Multi-NH-192}    & \textbf{192}   & \textbf{4816}  & \textbf{14406} &  \textbf{19222}   & \textbf{NO}    \\
\textbf{RQC.EGK-Multi-UR-192}             & \textbf{192}  & \textbf{2426} &  \textbf{9419} &  \textbf{11845} & \textbf{NO}                \\
LT19 \cite{ref35}  & 192 & 3193 & 3193 & 6386 & NO\\
Multi-RQC-AG-192 \cite{ref7} & 192 & 888 & 6780 & 7668  & YES  \\
NH-Multi-RQC-AG-192 \cite{ref7} & 192 & 979 & 6912 & 7891  & YES  \\
Multi-UR-AG-192 \cite{ref7} & 192 & 8375 &12700  &  21075 & YES  \\
NH-Multi-UR-AG-192 \cite{ref7} & 192 & 5133 & 7469 & 12602 & YES  \\
BW-RQC \cite{ref10} & 192 & 1834   & 3652  & 5486  & YES \\
EG-RQC \cite{song2025interleaved} & 192 & 1711   & 3342  & 5053  & YES \\
\hline\noalign{\smallskip}
HQC   \cite{ref20}           & 256      & 7245  &14485  & 21730  & YES                 \\
BIKE  \cite{ref21}            & 256      & 5122  & 5154 & 10276  & YES                \\
Classic McEliece \cite{ref22} & 256      & 1044992 & 208 & 1045200  & NO                 \\
LOI17  \cite{ref12}            & 256      & 59136  & 1920 & 61056  & NO                 \\
Classic RQC \cite{ref10} & 256 & 2421 & 4826 & 7247 & NO \\
\textbf{RQC.EGK-BWE-256} & \textbf{256} & \textbf{16002} & \textbf{31924}  & \textbf{47926} & \textbf{NO}                  \\ 
\textbf{RQC.EGK-Multi-NH-256}     & \textbf{256}      &  \textbf{6792}  & \textbf{37004}  & \textbf{43796}  & \textbf{NO}    \\
\textbf{RQC.EGK-Multi-UR-256}  & \textbf{256}      & \textbf{3831}  & \textbf{15269}   & \textbf{19100}   & \textbf{NO}                \\
LT19 \cite{ref35}  & 256 & 4291 & 4291 & 8582 & NO\\
BW-RQC \cite{ref10} & 256 & 2421   & 4826  & 7247  & YES \\
EG-RQC \cite{song2025interleaved} & 256 & 3190   & 6300  & 9490  & YES \\
\noalign{\smallskip}\hline
\end{tabular}
\end{table}

\section{Conclusion}
\label{sec888}

This paper has made three key contributions. First, further in-depth research has been conducted on the minimum rank distance of Gabidulin-Kronecker product codes, and results under two different parameter settings have been presented. In particular, under the condition of $n_{1}=k_{1}$, $n_{2}=m < n_{1}n_{2}$, the minimum rank distance of Gabidulin-Kronecker product codes has been shown to achieve the Singleton-type bound, thus, a new MRD code has been obtained. Second, for the first time, we have investigated the Extended Gabidulin codes with a Kronecker product structure and have proposed a new decoding method for Extended Gabidulin codes. Unlike some prior methods, our decoder directly recovers the codeword, ensuring zero decoding failure probability when the error weight is within the correction capability. Third, leveraging EGK codes, we constructed three enhanced variants of the RQC cryptosystem: RQC.EGK-BWE, RQC.EGK-Multi-NH, and RQC.EGK-Multi-UR. These schemes successfully integrate advanced techniques like blockwise errors, non-homogeneous errors, and unstructured matrices. A key achievement is that all variants maintain zero decryption failure probability while achieving competitive, and in many cases smaller, public key sizes compared to the classic RQC and other NIST round candidates.

Future work includes deriving exact formulas for the minimum distance of GK/EGK codes in the general parameter regime, which remains an open problem. Exploring other code compositions to construct new families of MRD codes is another promising direction. Finally, adapting the EGK code framework to design efficient signature schemes or other cryptographic primitives based on the rank metric merits investigation.


\end{document}